\documentclass[a4paper,11pt]{article}
\pdfoutput=0 % if your are submitting a pdflatex (i.e. if you have
\usepackage{jinstpub} % for details on the use of the package, please
\usepackage{color}% for personal comment

\title{\boldmath Towards Gotthard-II: Development of A Silicon Microstrip Detector for the European X-ray Free-Electron Laser}

\author[a,1]{Jiaguo Zhang,\note{Corresponding author.}}
\author[a]{Marie Andr\"a,}
\author[a]{Rebecca Barten,}
\author[a]{Anna Bergamaschi,}
\author[a]{Martin Br\"uckner,}
\author[a]{Roberto Dinapoli,}
\author[a]{Erik Froejdh,}
\author[a]{Dominic Greiffenberg,}
\author[a]{Carlos Lopez-Cuenca,}
\author[a]{Davide Mezza,}
\author[a]{Aldo Mozzanica,}
\author[b]{Marco Ramilli,}
\author[a]{Sophie Redford,}
\author[a]{Marie Ruat,}
\author[a]{Christian Ruder,}
\author[a]{Bernd Schmitt,}
\author[a]{Xintian Shi,}
\author[a]{Dhanya Thattil,}
\author[a]{Gemma Tinti,}
\author[b]{Monica Turcato,}
\author[a]{Seraphin Vetter}

\affiliation[a]{Paul Scherrer Institut,\\5232 Villigen, Switzerland}
\affiliation[b]{European X-ray Free-Electron Laser Facility GmbH,\\Holzkoppel 4, 22869 Schenefeld, Germany}

\emailAdd{jiaguo.zhang@psi.ch}

\abstract{Gotthard-II is a 1-D microstrip detector specifically developed for the European X-ray Free-Electron Laser. It will not only be used in energy dispersive experiments but also as a beam diagnostic tool with additional logic to generate veto signals for the other 2-D detectors. Gotthard-II makes use of a silicon microstrip sensor with a pitch of either 50 $\mu$m or \mbox{25 $\mu$m} and with 1280 or 2560 channels wire-bonded to adaptive gain switching readout chips. \mbox{Built-in} analog-to-digital converters and digital memories will be implemented in the readout chip for a continuous conversion and storage of frames for all bunches in the bunch train. The performance of analogue front-end prototypes of Gotthard has been investigated in this work. The results in terms of noise, conversion gain, dynamic range, obtained by means of infrared laser and X-rays, will be shown. In particular, the effects of the strip-to-strip coupling are studied in detail and it is found that the reduction of the coupling effects is one of the key factors for the development of the analogue front-end of Gotthard-II. %Finally, the new prototype design is proposed. In addition, a model has been developed to describe the position interpolation based on $\eta$ algorithm and compared to measurement. The error of position interpolation has been analyzed.
}

\keywords{Radiation-hard detectors; Instrumentation for FEL; X-ray detectors}

\begin{document}
\maketitle
\flushbottom

%\linenumbers

\section{Introduction}
\label{sec:intro}

The European X-ray Free-Electron Laser (XFEL.EU) \cite{XFELGmbH, Altarelli2006} has been constructed in the Hamburg/Schenefeld region and available for user experiments since the second half of 2017. It delivers extrashort, high intense X-ray pulses with a peak brilliance $\sim$ 8 orders of magnitude higher than any other synchrotron radiation source. The duration of each X-ray pulse is less than \mbox{100 fs}. The pulses are operated in bunch trains, each consisting of 2700 X-ray pulses with a separation of \mbox{220 ns}. The bunch trains are repeated with 10 Hz. The unique X-ray beam and its time structure pose the following challenges to detectors used at the XFEL.EU: A dynamic range of 0, 1, ..., 10$^{4}$ $\times$ \mbox{12.4 keV} photons, a frame rate of 4.5 MHz, and last but not least radiation hardness up to 1 GGy for 3 years of operation.

There are several detector development projects currently running for the XFEL.EU. AGIPD \cite{Graafsma2009, Xintian2010, Beat2011, Xintian2012}, LPD \cite{Koch2013, Hart2012} and DSSC \cite{Porro2012} are the 2-D pixel detectors for experiments at the XFEL.EU. All pixel detectors are expected to be commissioned in 2017 and 2018. In addition to the 2-D pixel detector systems, Gotthard-II, a 1-D microstrip detector, is specifically developed for the XFEL.EU, based on Gotthard-I but with improved functionality \cite{Aldo2011}. The Gotthard-II development started in 2015 and detectors will be commissioned in mid 2018 \cite{Monica2014}. %The Gotthard-II project is a collaboration between PSI and the XFEL.EU starting from 2015.

The Gotthard-II detector will be employed in the von Hamos spectrometers and Johann spectrometer for energy-dispersive experiments at the Femtosecond X-ray Experiments (FXE) beamline. In addition, it will be used as spectrum analyzer by the beam diagnostic group, as well as by the FXE, SPB\footnote{SPB: Single Particles, Clusters and Bio-molecules} and MID\footnote{MID: Materials Imaging and Dynamics} beamlines. The potential scientific applications include, but are not limited to: X-ray emission/absorption spectroscopy, hard X-ray high resolution single-shot spectrometry (HiREX), energy dispersive experiments, beam diagnostics, as well as veto generation for the other detectors \cite{Monica2014}. For more examples of potential scientific applications, refer to \cite{Alonso-Mori2012, Zhang2014, Van2016}.

\subsection{Requirements of Gotthard-II at the XFEL.EU}

The Gotthard-II detector has less readout channels but similar complexity compared to the other 2-D detectors for experiments at the XFEL.EU. In addition, it is the only detector capable of measuring all the bunches in a train. To perform proper scientific experiments, Gotthard-II needs to achieve a frame rate of 4.5 MHz to match the particular bunch structure, a dynamic range up to 10$^{4}$ 12.4 keV photons and single photon resolution\footnote{Radiation damage in Gotthard-II is not a problem, since the ASICs can be properly shielded and the silicon sensor will see considerable less dose compared to 2-D detectors whose focal plane faces to the XFEL beam.}. A detailed specification can be found in table~\ref{Table_spec}. 

\begin{table}[htbp]
\centering
\begin{tabular}{|c|c|c|c|}

\hline
\textbf{Parameter} & \textbf{Value} & \textbf{Unit} \\
\hline
energy range & 3 - 25 & keV \\
\hline
sensor thickness & 450 $\vert$ 320 & $\mu$m  \\
\hline
quantum efficiency & 83.5\% $\vert$ 72.2\% & @12.4 keV \\
\hline
sensitive area & 64 $\times$ 8 $\vert$ 64 $\times$ 6 & mm$^{2}$  \\
\hline
pitch of strip & 50 $\vert$ 25 & $\mu$m  \\
\hline
number of strips & 1280 $\vert$ 2560 &  \\
\hline
sensor material & silicon &  \\
\hline
dynamic range & 10$^{4}$ & 12.4 keV photons  \\
\hline
linearity & better than 1\% &  \\
\hline
point spread function & O(pitch) &  \\
\hline
integration time & $\geq$ 100 & ns \\
\hline
cooling & air or liquid &  \\
\hline
noise & $\sim$ 200 & $e^{-}$ r.m.s. \\
\hline
signal-to-noise ratio (SNR) & $\geq$10 & @12.4 keV  \\
\hline
photon sensitivity (no gain switching) & single photon & @ >3.5 keV (SNR > 5) \\
\hline
frame rate & $\geq$ 4.5 & MHz \\
\hline
readout time & < 99.4 & ms for 2700 frames \\
\hline
readout latency of hit info & < 220 & ns  \\
%\hline
%radiation tolerance & 1 & GGy within 3 years \\
\hline
vacuum compatibility & 10$^{-4}$ - 10$^{-5}$ & mbar \\
\hline

\end{tabular}
\caption{Specification of Gotthard-II detector of 50 and 25 $\mu$m pitches. 450 $\mu$m and 320 $\mu$m thick sensors will be used for pitches of 50 and 25 $\mu$m, respectively.}
\label{Table_spec}
\end{table}

%Due to the limit of readout speed, the 2-D detectors will store images due to X-ray pulses of a bunch train in the "on-chip" memories of the Application-Specific Integrated Circuits (ASICs) and read them out during the time interval of 99.4 ms between two bunch trains. In case analogue memories used, part of the charge stored in the memory cells may droop away before being read out. This issue has been investigated extensively inside AGIPD consortium and it is found that the droop of charge depends on temperature, irradiation dose and dose rate as well \cite{Julian2011, Julian2013, Julian20131}. Its distribution is also quite wide across the detector.  Operating the detector at -20$^{\circ}$C will be helpful to reduce, but not completely suppress the droop effect. In addition, due to the limit of pixel size, which is a trade-off between spatial resolution and number of frames stored in the ASIC, AGIPD and LPD can only record 352 and 512 images from the 2700 pulses in an XFEL bunch train. 

Gotthard-II is equipped with on-chip Analog-to-Digital Converters (ADCs) and Static Random-Access Memories (SRAM, digital memory) capable of storing 2700 images for all X-ray pulses in a bunch train: The analogue signals, after passing through a charge sensitive pre-amplifier and a Correlated-Double Sampling (CDS) stage, are digitized by the ADCs immediately and the digital values are stored in the SRAM. All the 2700 images are read out during the bunch train spacing of 99.4 ms. This approach has several advantages over the use of analogue memories to store signals from the CDS output, as implemented in \mbox{\textit{e.g.}} AGIPD. The immediate digitization of the signals removes the problem connected with the droop of charge in analogue memories and the consequent need to cool the detector to a very low temperature in order to reduce such effects \cite{Julian2011, Julian2013, Julian20131}. It moreover removes the complexity related to the analogue readout and off-chip digitization, which require great care and corresponding resources, to avoid signal degradation. The analogue memories would in addition be very large in size and suffer from an on-chip cross talk problem \cite{Davide2016}. Another important function of Gotthard-II is the generation of veto signals for 2-D detectors depending on the interaction between an XFEL pulse and investigated sample. Since the 2-D detectors have limited memories and are not able to record all images from the 2700 pulses per bunch train, with the veto signals generated by Gotthard-II, useless images of 2-D detectors can be discarded and the corresponding memories re-used. For this purpose, additional logic circuitry used to generate veto signals will be implemented into the final ASIC. This circuitry will provide a one-bit hit information per channel, not stored in the SRAM but read out immediately at a rate of \mbox{4.5 MHz}\footnote{The current design of Gotthard-I can achieve a frame rate of $\leq$ 1 MHz, and features no built-in ADC, digital memories and veto-generation logic.}. This information will then be used by the FPGA on the readout board to generate the veto signal.

\subsection{Development strategies}

%For the development of Gotthard-II ASICs, the analogue front-end electronics including pre-amplifier and CDS, and the ADC togehter with SRAM are being designed separately. The performance of each part will be verified independently and they will be integrated together in a final stage to form the full size ASIC. %For each part, similar actions are taken: First, the performance of existing prototypes will be verified; In case some issues observed, the issues will be identified and understood; Finally, based on the understanding, the design of the front-end and ADC will be improved.

The main building blocks of the Gotthard-II ASIC, namely the analogue front-end electronics including pre-amplifier and CDS, the ADC and the SRAM have been designed and implemented separately in Multi-Project Wafer (MPW) runs. Therefore, each block can have its performance assessed independently and is integrated in the full-size ASIC only in case of proven full functionality. A first prototype version of a complete channel made out of blocks not yet rated as "final grade", has already been sent for production. This will provide information about the functionality and the interactions of all the building blocks when interconnected to form a channel within the multi-channel prototype.

The Gotthard-I and Jungfrau \cite{Aldo2014} readout ASICs have been used as a basis for the development of the Gotthard-II analogue front-end. The paper will focus on the performance of the existing front-end prototypes of Gotthard fabricated in UMC-110 nm technology while the ADC and the SRAM will be discussed in a separate paper.
%For the ADC, some serious problems have been observed, for example the speed of comparator cannot achieve 20 MHz, the Sample and Hold stage (S \& H) shows large noise and charging up the DAC array is time-consuming and thus missing codes have been seen, which needs a completely new design. The front-end 

\section{The architecture of the analogue front-end prototypes}

\begin{figure}
\small
\centering
\caption{The architecture of the analogue front-end prototypes.}
\includegraphics[width=150mm]{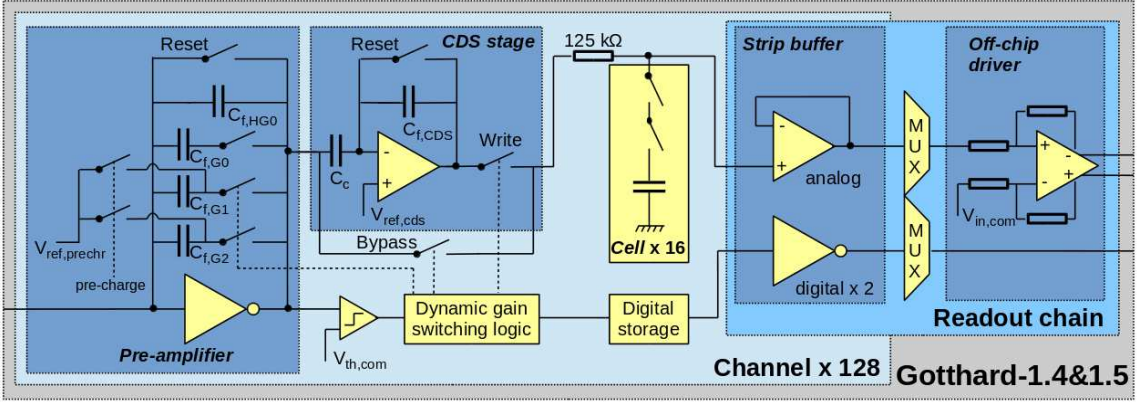}
\label{Architecture}
\end{figure}

The architecture of the analogue front-end prototypes (version Gotthard-1.4 \& -1.5) is shown in figure~\ref{Architecture}. It includes four main parts: 1) a dynamic gain switching pre-amplifier, 2) a CDS stage, 3) analogue and digital memory cells, and 4) a readout chain for all strip channels.

The pre-amplifier is a charge-sensitive pre-amplifier with dynamic gain switching functionality, similar to AGIPD \cite{Xintian2010} and Jungfrau \cite{Aldo2014}. Its output is connected to a comparator and a dynamic gain switching logic. There are four different feedback capacitors implemented in the pre-amplifier: $C_{f,HG0}$, $C_{f,G0}$, $C_{f,G1}$ and $C_{f,G2}$. Initially, either $C_{f,HG0}$ or $C_{f,HG0}+C_{f,G0}$ can be selected as feedback capacitance. During charge integration, if the output voltage moves above the threshold of the comparator, $V_{th,com}$, the dynamic gain switching logic will force the gain switching and the capacitor $C_{f,G1}$ will be added to the feedback loop of the pre-amplifier. This will cause a reduction of the pre-amplifier gain and, as a side effect, a charge redistribution and a consequent reduction of the output voltage of the  pre-amplifier. If the output voltage of the pre-amplifer is still above $V_{th,com}$, a second gain switching occurs by adding another feedback capacitor, $C_{f,G2}$, to the feedback circuit. $C_{f,G1}$ and $C_{f,G2}$ can be pre-charged during the pre-amplifier reset phase. In this way, the output voltage range of the pre-amplifier after gain switching can be maximized, thus a larger dynamic range can be achieved. For the convenience, in the following we will note the gain using $C_{f,HG0}$ as HG0, with $C_{f,G0}$, $C_{f,G1}$ and $C_{f,G2}$ in addition as G0, G1 and G2, respectively. %Both HG0 and G0 can achieve single photon resolution at 8.05 keV but show different noise performance, which will be shown in Section~\ref{subsec:gain} and \ref{subsec:noise}. 

%A comparator is connected to the output of the pre-amplifier. As mentioned before, if the output voltage is below the threshold, $V_{th,com}$, during charge integration, the dynamic gain switching block will force the gain switching by means of the capacitor $C_{f,G1}$ will be added to the feedback loop of the pre-amplifier. This will cause a charge redistribution and thus reduce the output voltage of the pre-amplifier. If the output voltage of the pre-amplifer is still below $V_{th,com}$, a second gain switching occurs by adding another feedback capacitor, $C_{f,G2}$, to the feedback circuit. The digital memory of 2-bit depth of each channel stores the information indicating which gain is used.

The CDS stage is connected to the output of the dynamic gain switching pre-amplifier. It is used to remove the low frequency noise and the pre-amplifier reset noise. The amplification factor of the CDS is 2.35 (also called "CDS gain"). If gain switching happens, the correlation of the initial sample, stored in the CDS circuitry, and the actual signal is lost, so that CDS is not beneficial any longer. For this reason, the CDS stage is bypassed after gain switching. The signal is written into the analogue memory cells through a resistor of 125 k$\Omega$ which, together with the capacitive load at the CDS output, is used as an additional low-pass filter for noise reduction.

The analogue signals from the CDS output are stored in analogue memory cells, while the information indicating the gain is stored, for each channel, in a 2-bit digital memory.%while the digital memories of 2-bit depth of each channel stores the information indicating which gain is used.

%The readout chain includes an analogue and a digital buffer for each channel and an off-chip driver. 
During read-out, analogue and digital storage cells are driven by analogue and digital buffers separately. The analogue signals are selected by a multiplexer (MUX) and converted to fully differential signals through an off-chip driver and finally digitized by 14-bit ADCs on the readout board; The digital signals are sampled by a Field Programmable Gate Array (FPGA) on the readout board directly.

The investigated Gotthard-1.4 \& -1.5 prototype ASICs are wire-bonded to \mbox{320 $\mu$m} thick silicon micro-strip sensors with 128 strips of 50 $\mu$m pitch and 8 mm length for testing. The only difference between Gotthard-1.4 and -1.5 ASICs is the size of the transistors used in the pre-amplifier, which is supposed to influence the speed and the noise of the pre-amplifier. Since the speed of writing charge into the analogue memory cell is limited by the serial resistor in the circuit, the difference in the speed of the pre-amplifier between the two prototypes cannot be measured. Thus, only results from Gotthard-1.5 will be shown and discussed in Section \ref{sec:performance}.

%\subsection{Writing into memory}

%The current at the input node of the pre-amplifier will be integrated and the charge stored on the feedback capacitor changes the voltage at the pre-amplifier output. In HG0 and G0, the output voltage of the pre-amplifier will be sampled twice by the CDS stage: once before photons enter and once before the end of the integration. The voltage difference between the two samplings will be stored into the analogue memory cells. For G1 and G2 after gain switching, the output voltage of the pre-amplifier will be written into the analogue memory cells directly.

%\subsection{Read-out}

%During reading out, analogue memory and digital storage will be driven by analogue and digital buffers of each strip. In addition, the analogue signals are shifted out by a multiplexer (MUX) and differentiated by an off-chip driver. The output signal of the ASIC is converted by 14-bit ADCs on the readout board. 

\section{The performance of the prototypes}
\label{sec:performance}

The performance of the front-end prototypes in terms of conversion gain, noise, dynamic range and strip-to-strip coupling has been investigated experimentally. All measurements were performed at room temperature and the prototype assemblies were cooled by a fan. The sensor was biased at \mbox{240 V} and the power supply voltage of the ASICs was 1.4 V. %The threshold voltage, $V_{th,com}$, of the comparator coupled to the output the pre-amplifier to trigger the dynamic gain switching block (as seen in figure~\ref{Architecture}), was 330 mV and the medium and low gain capacitors of the pre-amplifier $C_{f,G1}$ and $C_{f,G2}$ were pre-charged to 940 mV by $V_{ref,prechr}$. \textcolor{red}{Shall we include these details?}

\subsection{Conversion gain}
\label{subsec:gain}

%The conversion gain in the text refers to the gain for HG0 and G0 with a unit of ADU/keV. 

In the following discussion, the conversion gain refers to the gain for HG0 and G0 and is expressed in ADU/keV. The conversion gain is determined using X-ray fluorescence of copper (Cu), which was placed as the target of an X-ray tube. The characteristic energy of the main $k_{\alpha}$ line of the fluorescence, $E_{k_{\alpha}}$, is 8.05 keV. In the measurement, an integration time of 10 $\mu$s was used and 100k frames were collected.

\begin{figure}
\small
\centering
\caption{The conversion gain of the prototypes measured with an integration time of 10 $\mu$s. (a) Fitting to multiple peaks due to 0, 1, 2 and 3 photons for a single strip (strip-64) as an example; (b) Linear fit to the peak positions to determine the conversion gain; (c) Conversion gain for all channels for HG0 and G0 following the same procedure; (d) Histogram of conversion gain for HG0 and G0.}
\vskip 0.2in
\includegraphics[width=75mm]{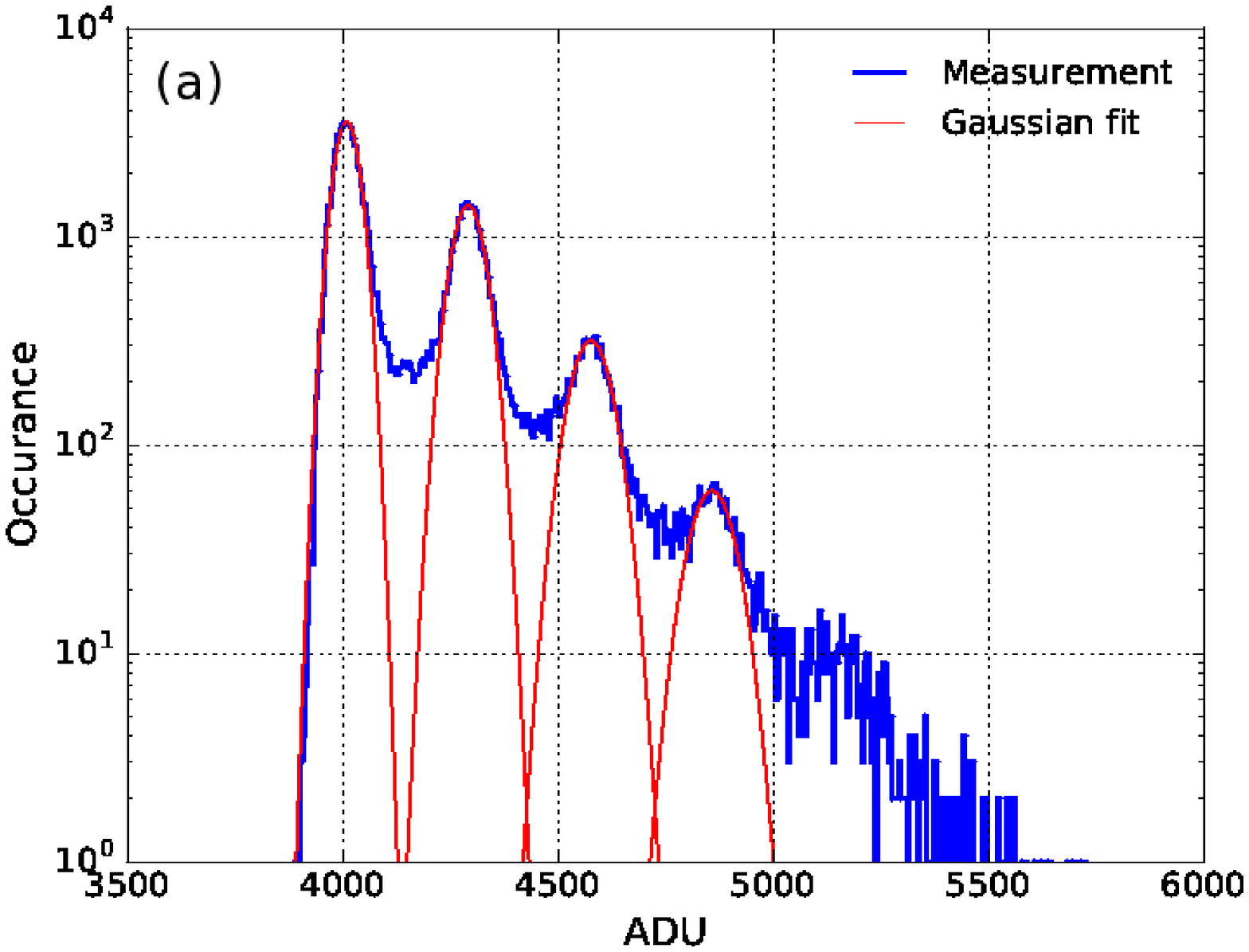}
\includegraphics[width=75mm]{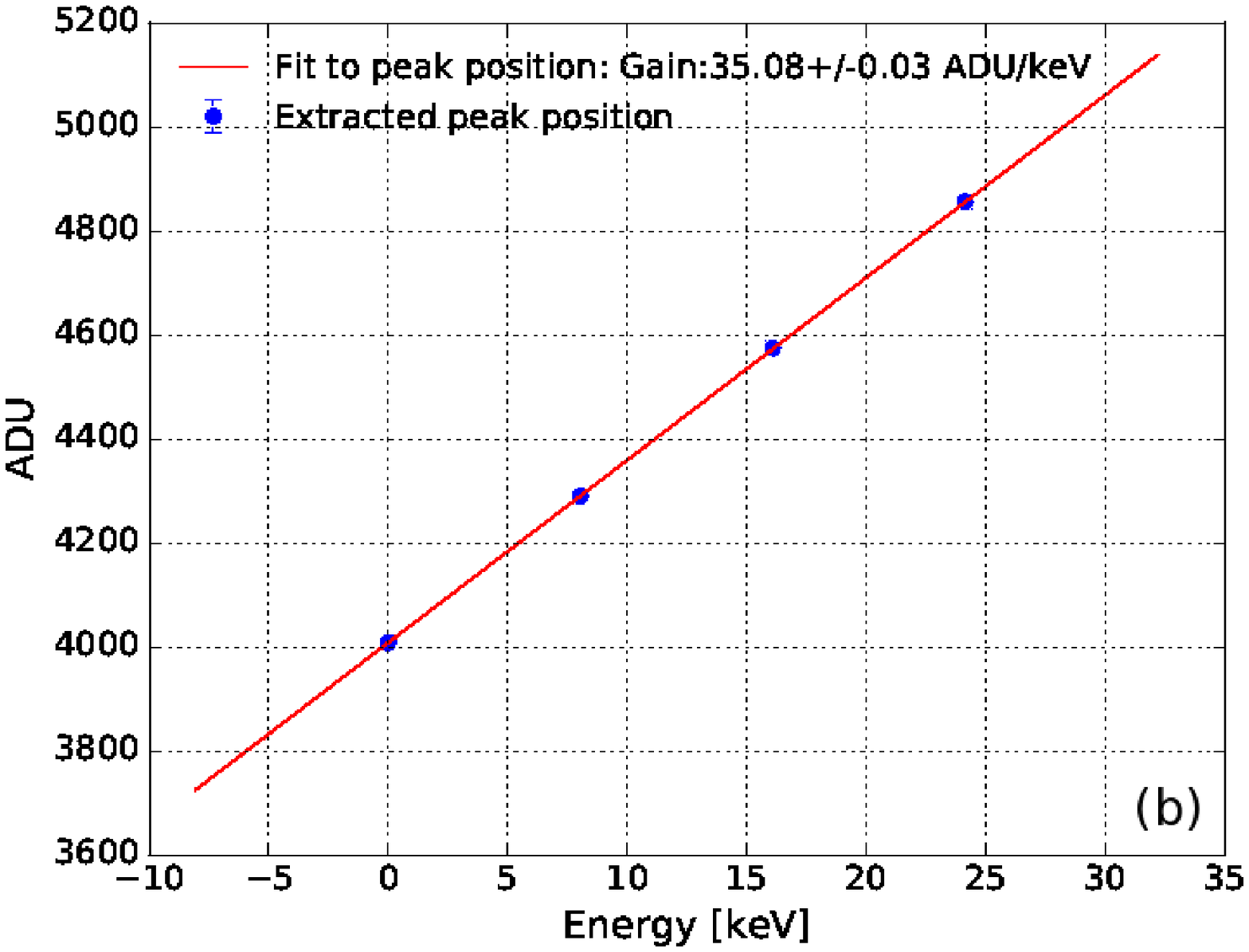}
\vskip 0.2in
\includegraphics[width=75mm]{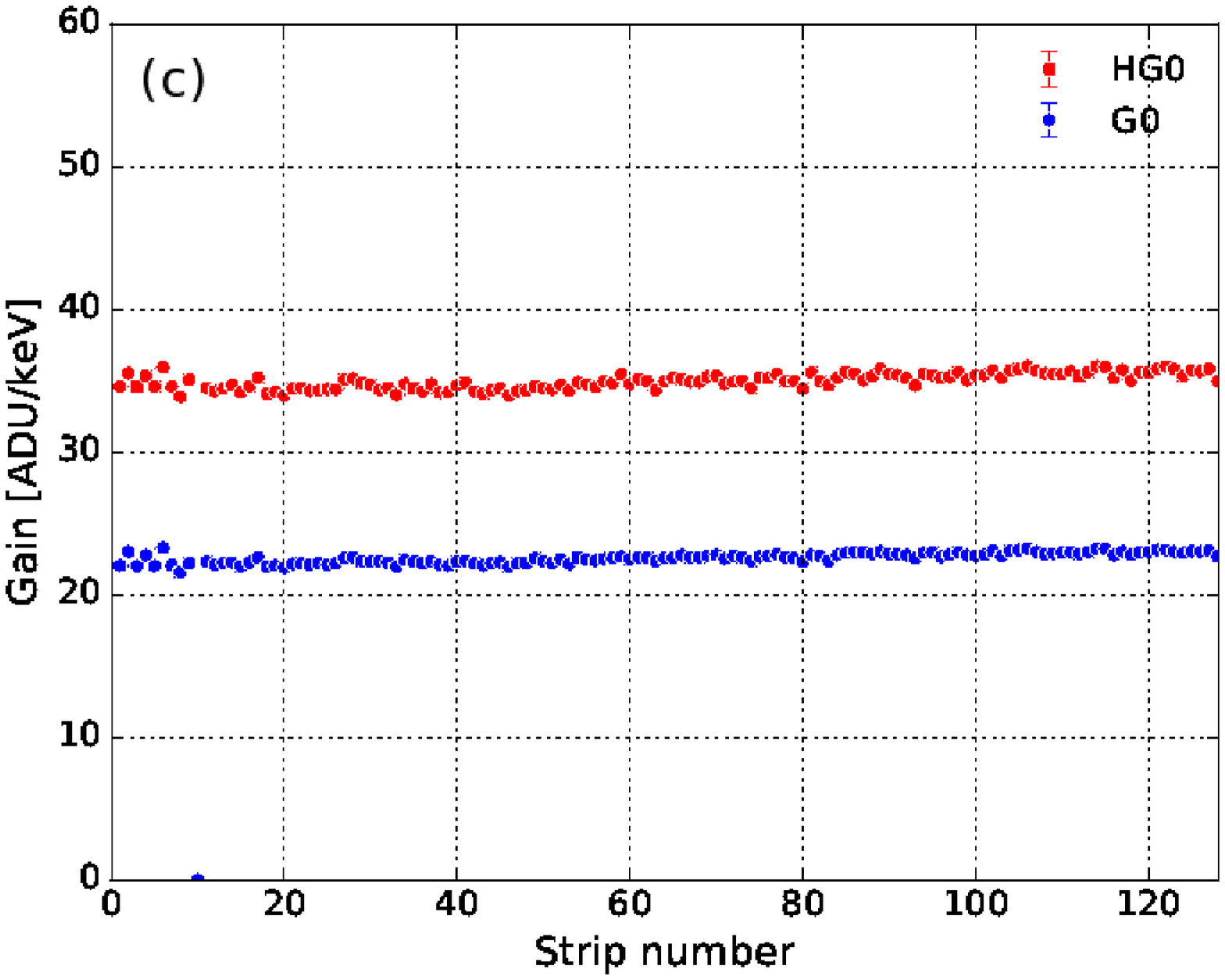}
\includegraphics[width=75mm]{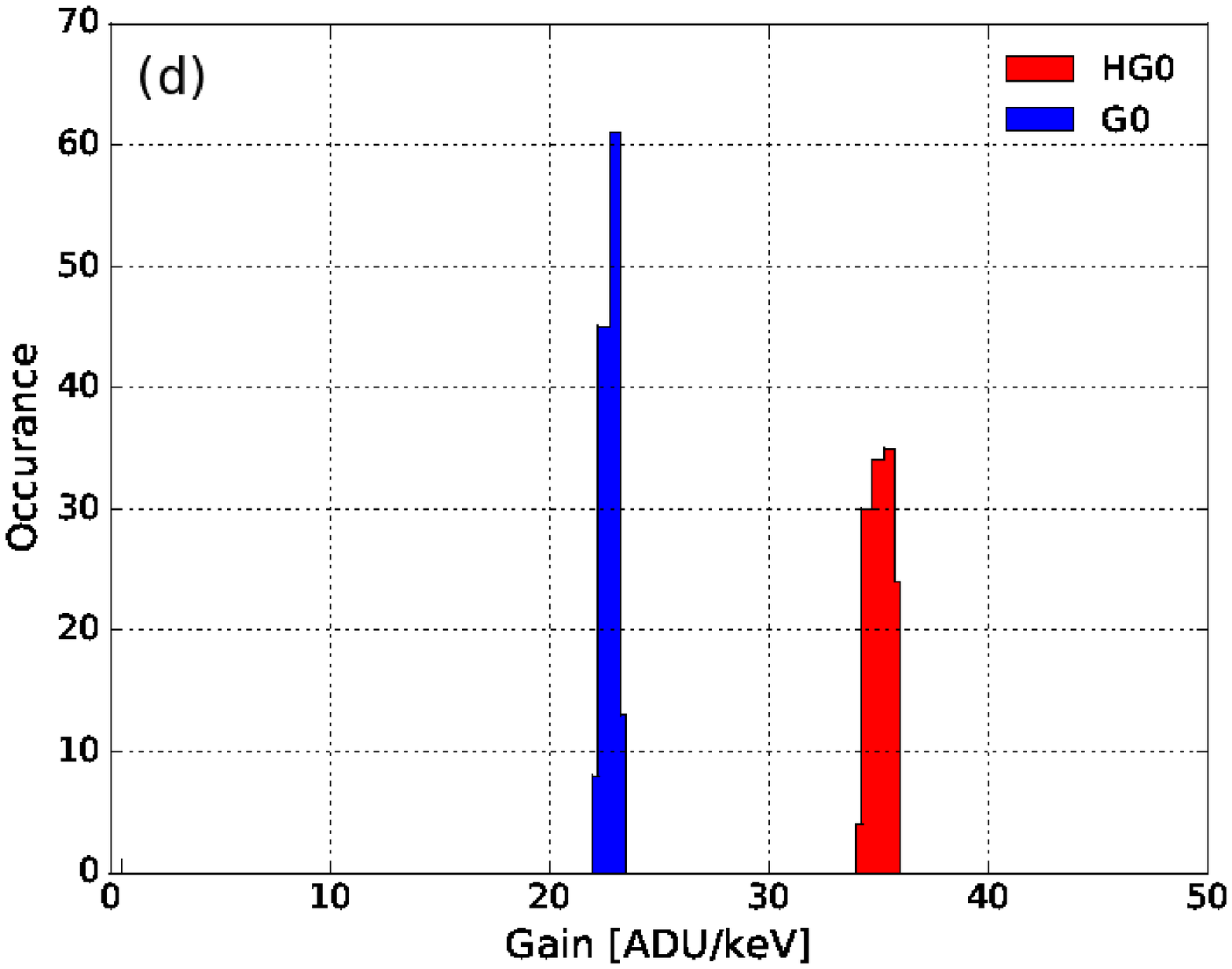}
\label{Conversion_gain}
\end{figure}

Figure~\ref{Conversion_gain}(a) shows the histogram of the measured ADU values using HG0 for a specific channel as an example. The identified peaks in the figure refer to 0, 1, 2 and 3 photons. Good separation between different photon peaks can be seen, indicating good noise performance. The peak positions were extracted from a Gaussian fit to each individual peak. Figure ~\ref{Conversion_gain}(b) shows the extracted peak position in terms of ADU as function of energy, which is given by the energy of the $k_{\alpha}$ X-ray fluorescence times number of photons. The slope of a linear fit gives the conversion gain. It should be noted that the intensity of the $k_{\beta}$-line at 8.90 keV of the cooper foil is much lower than the $k_{\alpha}$-line and cannot be resolved in the distribution due to the influence of noise as well as the charge diffusion, thus has been neglected in our case.

Figure~\ref{Conversion_gain}(c) shows the conversion gains of all channels for HG0 and G0, and figure~\ref{Conversion_gain}(d) the histogram of the gain distributions. The conversion gains for HG0 and G0 are centered at \mbox{35.2 $\pm$ 0.8 ADU/keV} and 22.8 $\pm$ 0.4 ADU/keV with $\sim$ 2\% channel to channel variations, which shows very good uniformity over all channels of the ASIC. 

By means of the ratio of the conversion gain between HG0 and G0, the parasitic capacitance in the feedback loop of the pre-amplifier can be estimated to be 40.5 fF.

\subsection{Noise}
\label{subsec:noise}

The noise measurement was performed in a light-tight box by measuring the integrated leakage current of the sensor for 10 $\mu$s multiple times. The histogram of ADU values was then fitted by a Gaussian function and the standard deviation, $\sigma$, was extracted. The noise is obtained using:

\begin{equation}
\label{eq:noise}
  noise\ r.m.s. [e^{-}] = \frac{\sigma [ADU]}{gain [ADU/keV]} \cdot \frac{1000}{3.6 [eV]}
\end{equation}

\noindent where 3.6[\textit{eV}] is the mean energy needed to generate one electron-hole pair in silicon by ionizing radiation, and \textit{gain}[\textit{ADU/keV}] is the previously measured conversion gain.

\begin{figure}
\small
\centering
\caption{The noise of the prototypes measured with an integration time of 10 $\mu$s for HG0 and G0. (a) Noise for all channels; (b) Histogram of the noise distribution.}
\vskip 0.2in
\includegraphics[width=75mm]{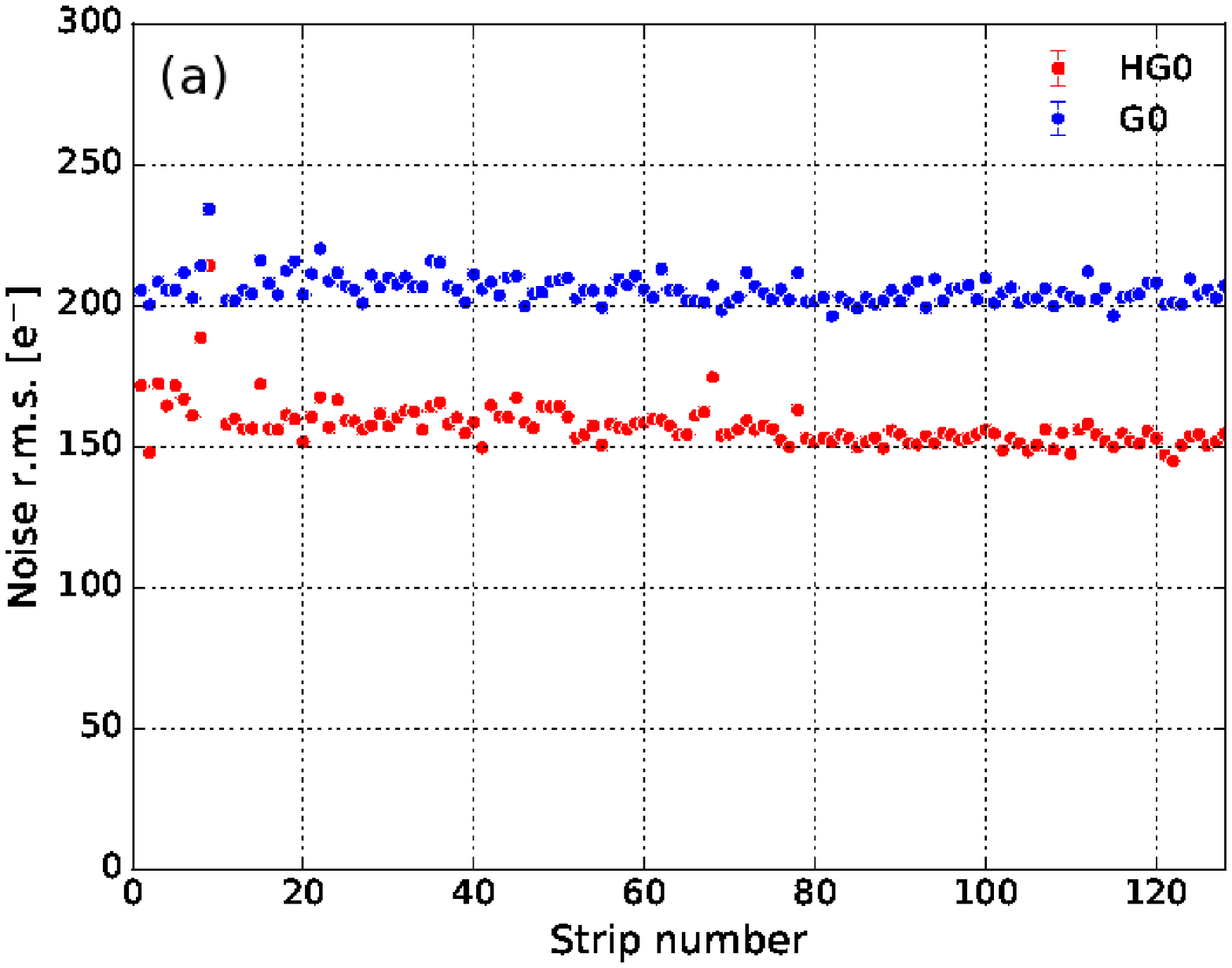}
\includegraphics[width=75mm]{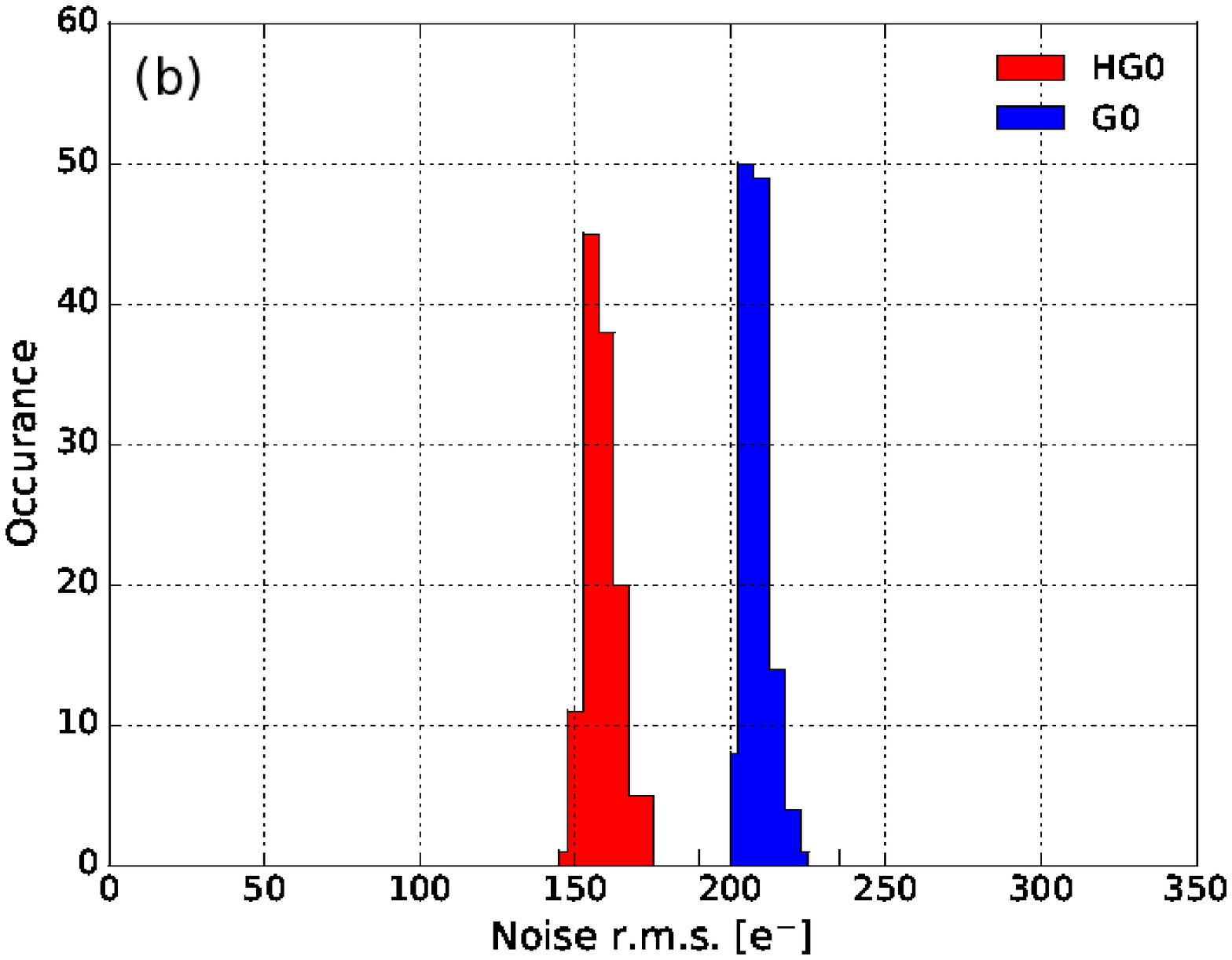}
\label{Noise_extraction}
\end{figure}

\begin{figure}
\small
\centering
\caption{The noise, derived from the extracted $\sigma$ from Gaussian fit, as function of integration time for all strip channels.}%; (b) The extrapolated noise over all channels at 100 ns; (c) Histogram of the extrapolated noise distributions at 100 ns.}
\vskip 0.2in
\includegraphics[width=120mm]{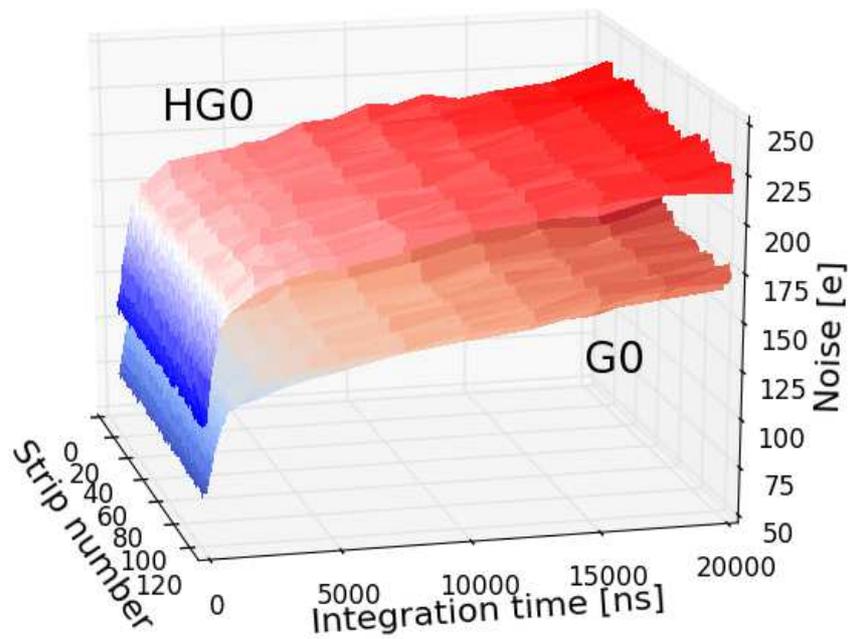}
\label{Noise_extrapolation}
\end{figure}

The noises for HG0 and G0 are $158 \pm 5\ e^{-}$ and $208 \pm 4\ e^{-}$ for an integration time of 10 $\mu$s, as shown in figure~\ref{Noise_extraction}. In addition, the noise has been investigated for different integration times, from \mbox{20 $\mu$s} down to 50 ns. Figure~\ref{Noise_extrapolation} shows the extracted noise as function of integration time for all strip channels: The conversion gain obtained from the X-ray fluorescence measurement with an integration time of 10 $\mu$s was applied to the extracted $\sigma$ at different integration times using formula~\ref{eq:noise}. A reduction of $\sigma$ below 500-600 ns has been observed which is due to the $RC$ time constant in the circuit: The 125 k$\Omega$ resistor, that in Jungfrau helps to remove the high frequency noise, limits the writing speed. When writing charge from the CDS output into analogue memories, at least \mbox{500-600 ns} are needed for the signal to settle.% To explore the noise corresponding to the integration time at the XFEL.EU ($\sim$ 100 ns), the curve above 600 ns was fitted by: 

%\begin{equation}
%\label{eq:noise_int}
 % noise\ r.m.s. [e^{-}]=\sqrt{k \cdot t_{int} + b^{2}}
%\end{equation}

%\noindent with $t_{int}$ the integration time and $b$ the intrisic noise at $t_{int}=0$. %Figure~\ref{Noise_extrapolation}(b) and (c) show the noise extrapolated to the XFEL.EU timing and its distribution over all channels: A noise of $129 \pm 5\ e^{-}$ and $180 \pm 4\ e^{-}$ for HG0 and G0, which fulfills the noise requirement for Gotthard-II, has been obtained after extrapolation.

\subsection{Dynamic range}
\label{subsec:DR}

The dynamic range was measured using a pulsed infrared laser with a wavelength of 1030 nm. The integration time was set to 5 $\mu$s and the pulse duration was less than one nano second. The dynamic range scan was done by varying the laser intensity. The laser intensities were calibrated by measuring the photo-current from a 320 $\mu$m thick planar silicon diode and then converting to equivalent number of \mbox{12.4 keV} photons per pulse. A detailed introduction to the experimental setup and conversion from laser intensity to photons can be found in \cite{Dominic2014, Davide2016}.

\begin{figure}
\small
\centering
\caption{Dynamic range scan with infrared laser using HG0 (in blue) and G0 (in green). (a) Dynamic range scan; (b) Shot-to-shot laser fluctuation.}
\vskip 0.2in
\includegraphics[width=75mm]{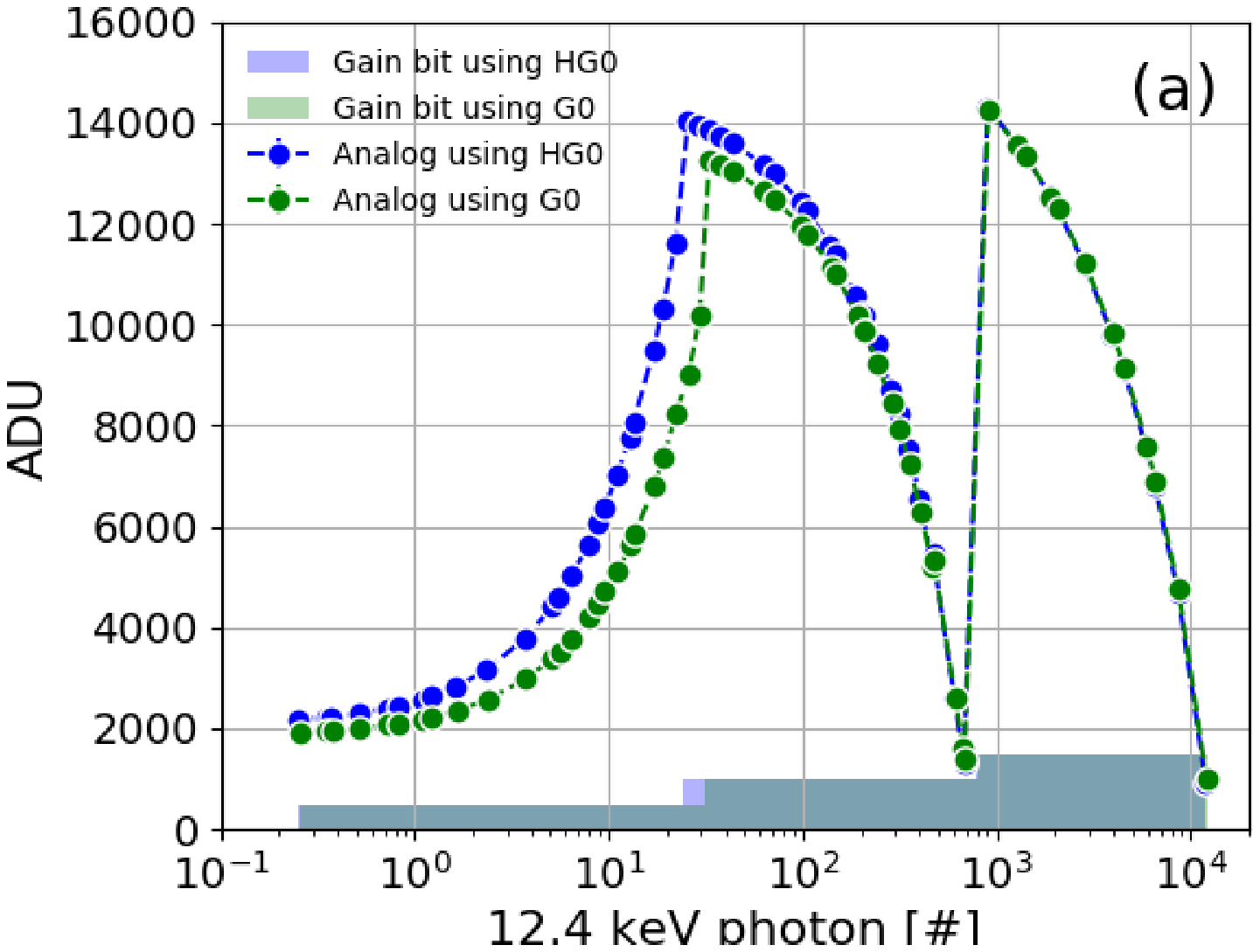}
\includegraphics[width=75mm]{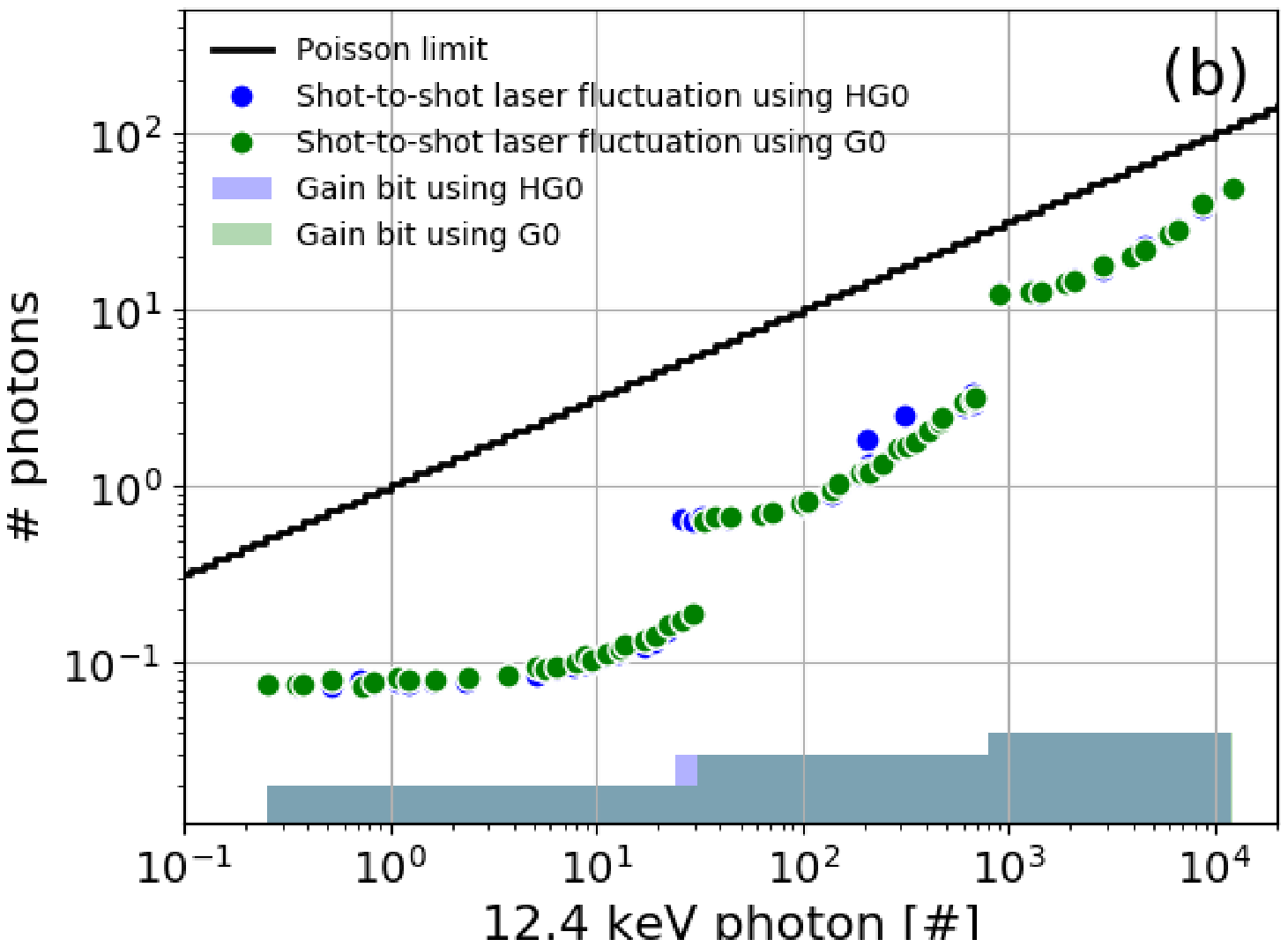}
\label{DRscan}
\end{figure}

Figure~\ref{DRscan}(a) shows the dynamic range for a specific channel using HG0 and G0. It can be seen the dynamic range is up to $1.26 \times 10^{4}$ 12.4 keV photons at the end of the scan. The ratio of gains, obtained from the ratio of the slopes of the linear fits to the measurement points in different gains, has been summarized in table~\ref{Table_ratio}. The increase of the ratio at higher intensites in all gains in figure~\ref{DRscan}(b) comes from the shot-to-shot fluctuation in the laser intensity, which results in a convolution of the laser fluctuations and the electronic noise of the ASIC. The fluctuations are well below the Poisson limit over the entire dynamic range. Considering $5\sigma$ as a good separation to resolve single photons, in HG0 and G0 single photon resolution can be achieved for X-ray photons with an energy above 3.7 keV; in G1 and G2, it is possible to resolve 3 and 55 photons of 12.4 keV, respectively.

\begin{table}[htbp]
\centering
\begin{tabular}{|c|c|c|c|c|}

\hline
\textbf{Gain ratio} & HG0/G1 & HG0/G2 & G0/G1 & G0/G2 \\
\hline
\textbf{Measured value} & $22.11 \pm 0.08$ & $346.1 \pm 1.4$ &  $15.47 \pm 0.05$ & $230.2 \pm 0.9$ \\
\hline

\end{tabular}
\caption{The ratio of gains in different gain stages using HG0 or G0.}
\label{Table_ratio}
\end{table}

\subsection{Strip-to-strip coupling}

Due to the capacitive coupling between strip channels, even if all charge carriers generated by X-ray photons (no charge sharing effect) are collected by a strip, the neighbouring readout channels of the strip still measure a signal (also known as capacitive "charge division"). There are various models which describe this effect caused by the capacitive coupling between strip channels \cite{England1981, Lutz1991, Spieler2005, Turchetta1993, Dabrowski1994, Dabrowski1996, Krammer1997}. The charge measured by the strip channel collecting all carreries produced by X-ray photons, $Q^{i}$, the charge measured by one of its first neighbouring channels, $Q^{i+1}$, by its second and third neighbouring channels, $Q^{i+2}$ and $Q^{i+3}$, can be simplified as\footnote{Channel definition can be referred to figure~\ref{Coupling_sim_layout}}

\begin{equation}
\label{eq:Qm}
  Q^{i} = \dfrac{(A+1)\cdot (C_{f} + C_{para})}{(A+1)\cdot (C_{f} + C_{para}) + C_{inp}} \cdot Q^{tot}
\end{equation}

\begin{equation}
\label{eq:Q1st}
  Q^{i+1} = \dfrac{C_{c}^{1st}}{(A+1)\cdot (C_{f} + C_{para}) + C_{inp}} \cdot Q^{tot}
\end{equation}

\begin{equation}
\label{eq:Q2nd}
  Q^{i+2} = \dfrac{C_{c}^{2nd}}{(A+1)\cdot (C_{f} + C_{para}) + C_{inp}} \cdot Q^{tot}
\end{equation}

\begin{equation}
\label{eq:Q3rd}
  Q^{i+3} = \dfrac{C_{c}^{3rd}}{(A+1)\cdot (C_{f} + C_{para}) + C_{inp}} \cdot Q^{tot}
\end{equation}

\noindent with the assumption that $C_{c}^{1st,2nd,3rd} << (A+1)\cdot (C_{f}+C_{para})$. \textit{A} is the DC gain of the pre-amplifier, also known as open loop gain. $C_{c}^{1st}$, $C_{c}^{2nd}$ and $C_{c}^{3rd}$ are the coupling capacitances between the strip channel collecting all carriers and its first, second and third neighbours. The coupling capacitance includes the contributions from the interstrip capacitance of the silicon sensor, the coupling capacitances between bonding wires, as well as between bonding pads. $C_{f}$ is the feedback capacitance of the pre-amplifier of strip-$i$, $C_{para}$ the parasitic capacitance adding to the same feedback loop, $Q^{tot}$ the total charge, and $C_{inp}$ the total capacitance at the input node of strip-$i$. $C_{inp}$ is obtained by

\begin{equation}
\label{eq:Cinp}
  C_{inp} = C_{strip} + 2 \sum_{i} C_{c}^{i}
\end{equation}

\noindent with $C_{strip}$ the bulk capacitance of an individual strip. If we only consider the capacitive coupling up to the third neighbouring channel, formula~\ref{eq:Cinp} can be written as:

\begin{equation}
\label{eq:Cinp_simp}
  C_{inp} = C_{strip} + 2 \cdot (C_{c}^{1st} + C_{c}^{2nd} + C_{c}^{3rd})
\end{equation}

Thus, the coupling factor $k_{factor}$, defined by the ratio of charge collected by the neighbouring channel and the channel collecting the majority of the charge, is given by:

\begin{equation}
\label{eq:k_1st}
  k_{factor}^{1st} = \frac{Q^{i+1}}{Q^{i}} = \frac{C_{c}^{1st}}{(A+1)\cdot (C_{f} + C_{para})}
\end{equation}

\begin{equation}
\label{eq:k_2nd}
  k_{factor}^{2nd} = \frac{Q^{i+2}}{Q^{i}} = \frac{C_{c}^{2nd}}{(A+1)\cdot (C_{f} + C_{para})}
\end{equation}

\begin{equation}
\label{eq:k_3rd}
  k_{factor}^{3rd} = \frac{Q^{i+3}}{Q^{i}} = \frac{C_{c}^{3rd}}{(A+1)\cdot (C_{f} + C_{para})}
\end{equation}

Since $C_{strip} << 2 \cdot (C_{c}^{1st} + C_{c}^{2nd} + C_{c}^{3rd})$, the charge lost to the strip capacitance coupled to the backside is negligible \cite{Paolo1990, Ikeda1993}. In this case, the fractional charge measured by each strip channel can be calculated by

\begin{equation}
\label{eq:Q_frac}
  Q_{frac}^{i} = \frac{k_{factor}^{i}}{1 + 2 \sum_{i} k_{factor}^{i} }
\end{equation}

\noindent after the coupling factors have been determined.

The strip-to-strip coupling will be discussed in two cases: 1) Coupling before gain switching (all channels are in the same gain), and 2) coupling right after dynamic gain switching (channels not in the same gain).

\subsubsection{Coupling before gain switching}

To determine the coupling factor before gain switching, low-rate X-ray measurements (only 0 or 1 photon collected by each strip per frame) were performed. This can be done either by reducing the current of the X-ray tube or by decreasing the integration time. Since the fractional charge in the neighbouring channels is of the same order as the noise charge, the determination has to be based on a large statistic with enough photon entries.

\begin{figure}
\small
\centering
\caption{2D map of charge measured by strip-$i$ vs. strip-($i+1$), strip-($i+2$) and strip-($i+3$) in HG0. In this figure, strip-$40$ ($i = 40$) and its three neighbouring strips were used to generate the plot.}
\vskip 0.2in
\includegraphics[width=140mm]{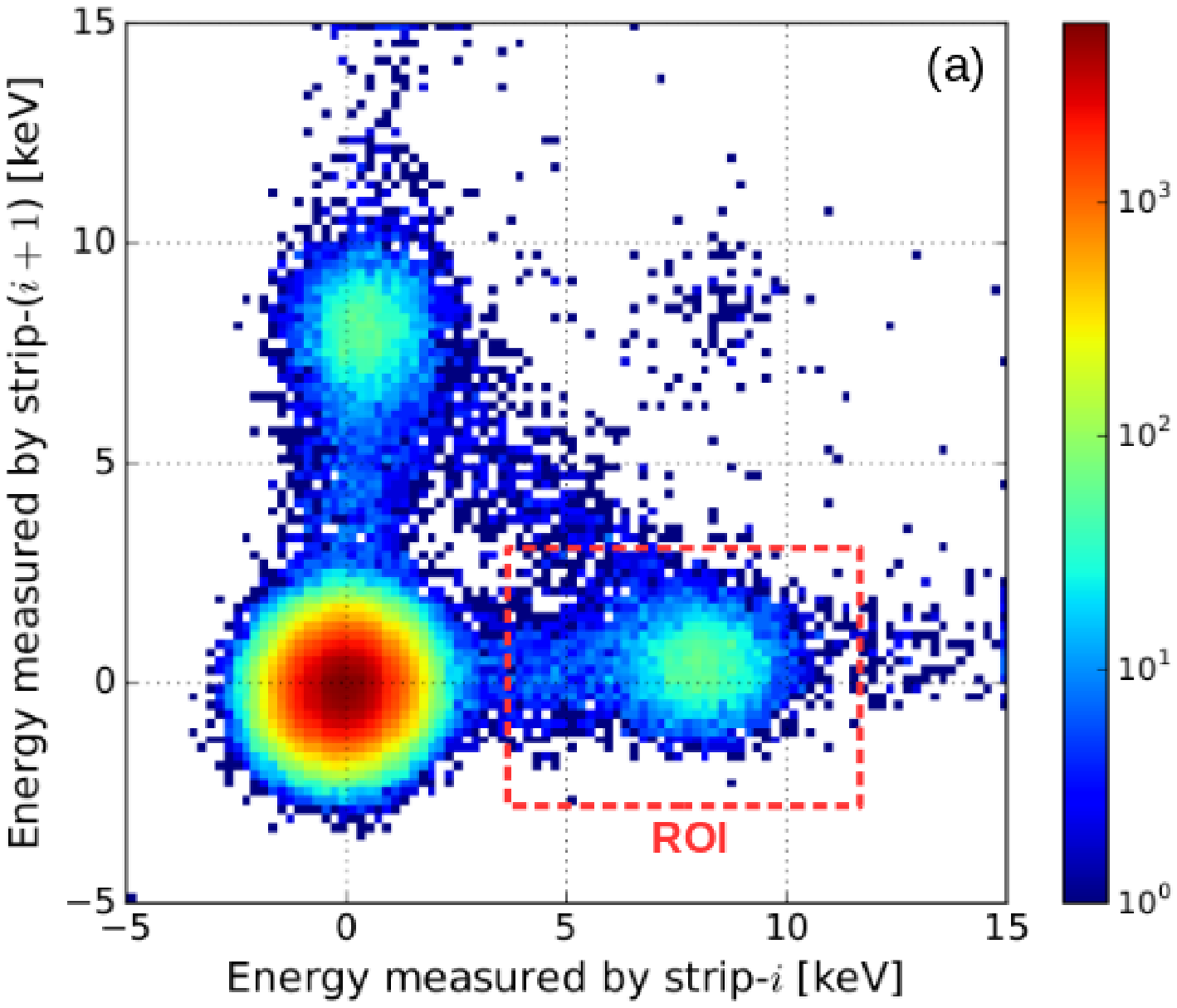}
\vskip 0.2in
\includegraphics[width=75mm]{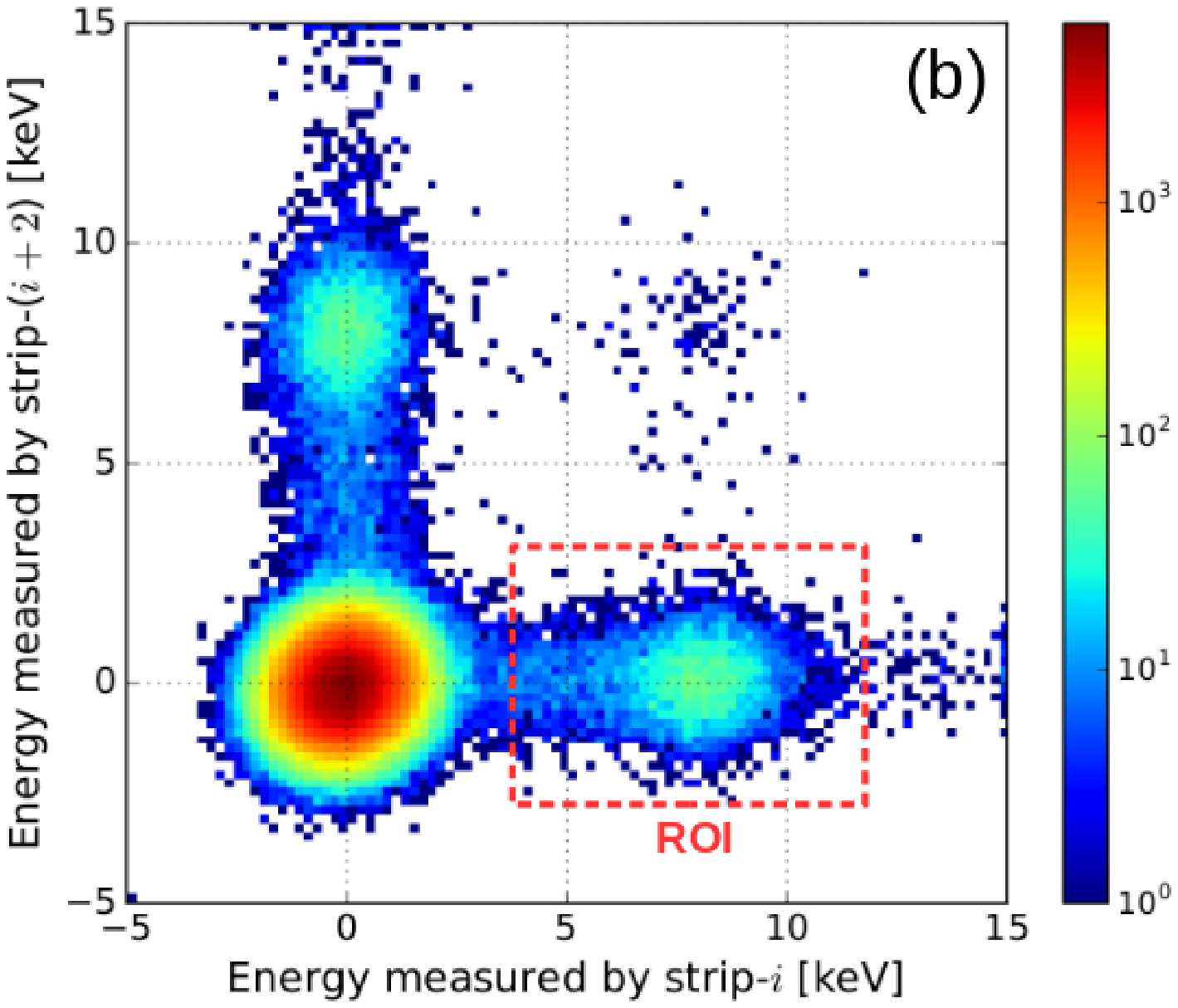}
\includegraphics[width=75mm]{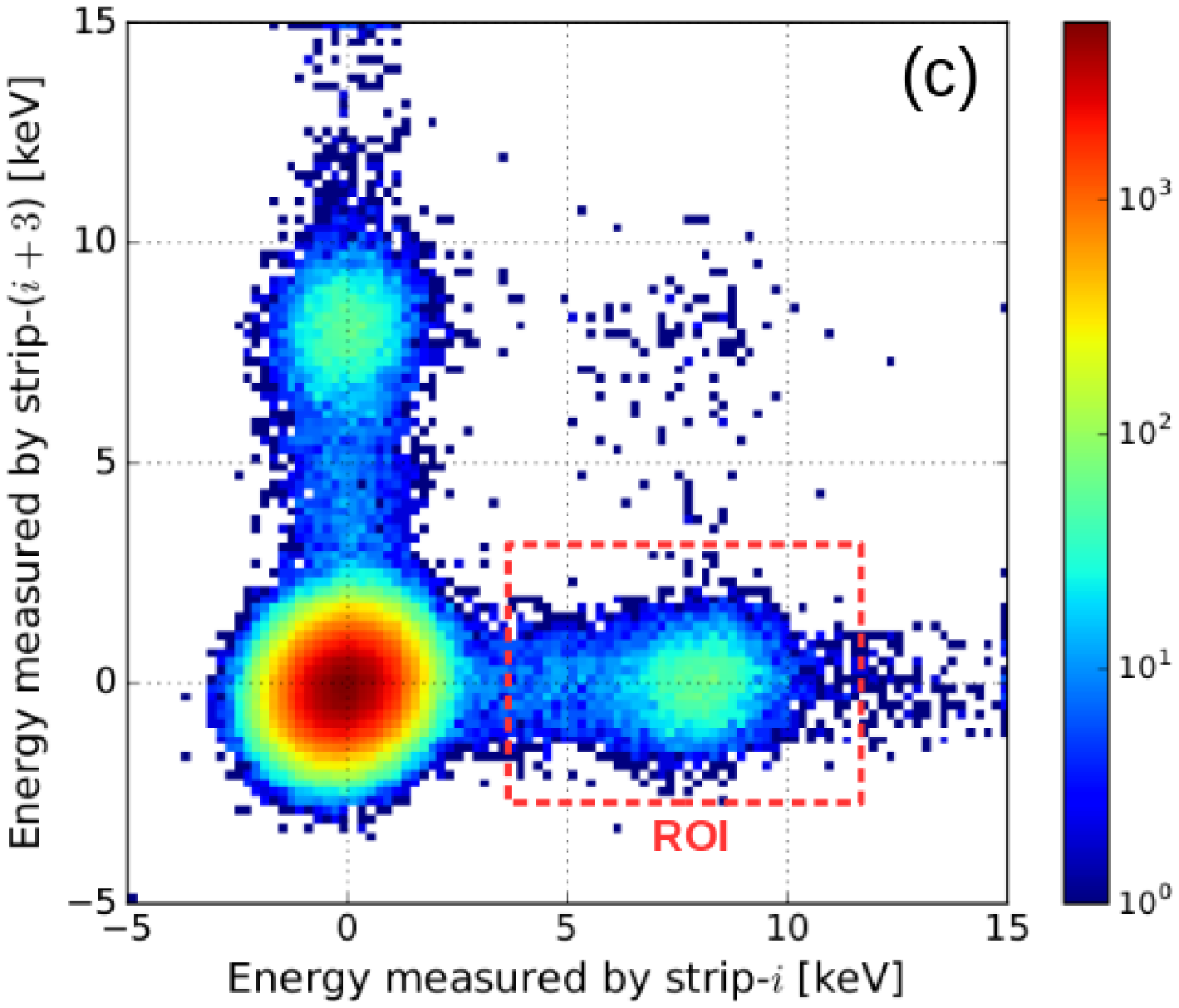}
\label{Coupling_2D}
\end{figure}

Figure~\ref{Coupling_2D} shows the relation between the energy measured by strip-$i$ and its first, second and third neighbouring channels on one side (noted as strip-($i+1$), strip-($i+2$) and strip-($i+3$)) in HG0\footnote{Results in G0 have also been obtained but are not shown here.}. The raw measurement in ADU values have been converted to energy based on the conversion gain determined from the X-ray fluorescence measurement.

\begin{figure}
\small
\centering
\caption{Projection of ROI onto the axes of strip-$i$, strip-($i+1$), strip-($i+2$) and strip-($i+3$): (a) in HG0; (b) in G0. In this figure, strip-$40$ ($i = 40$) and its three neighbouring strips were used to generate the plot. The coupling to strip-($i+n$) are clearly visible as a peak shift.}
\includegraphics[width=75mm]{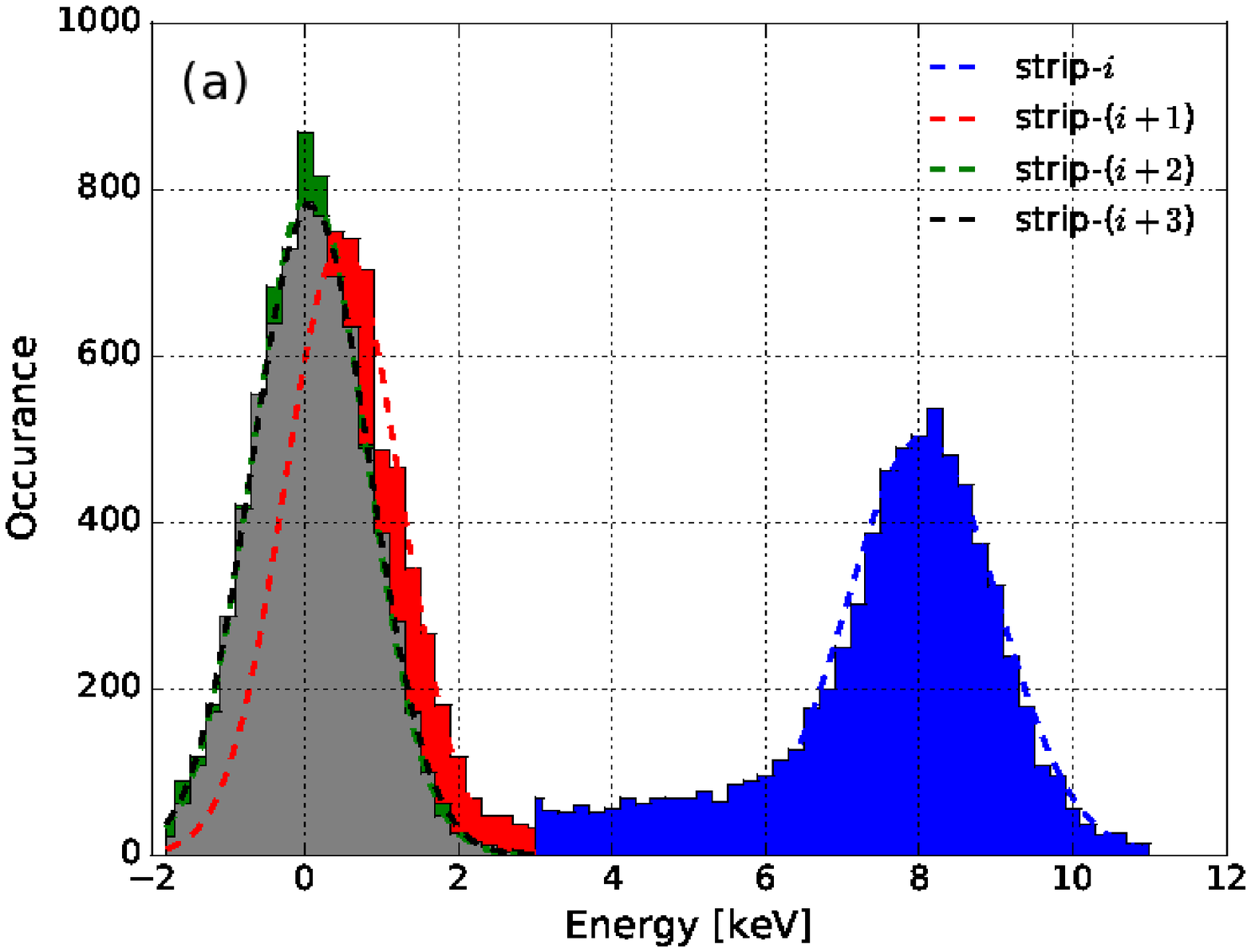}
\includegraphics[width=75mm]{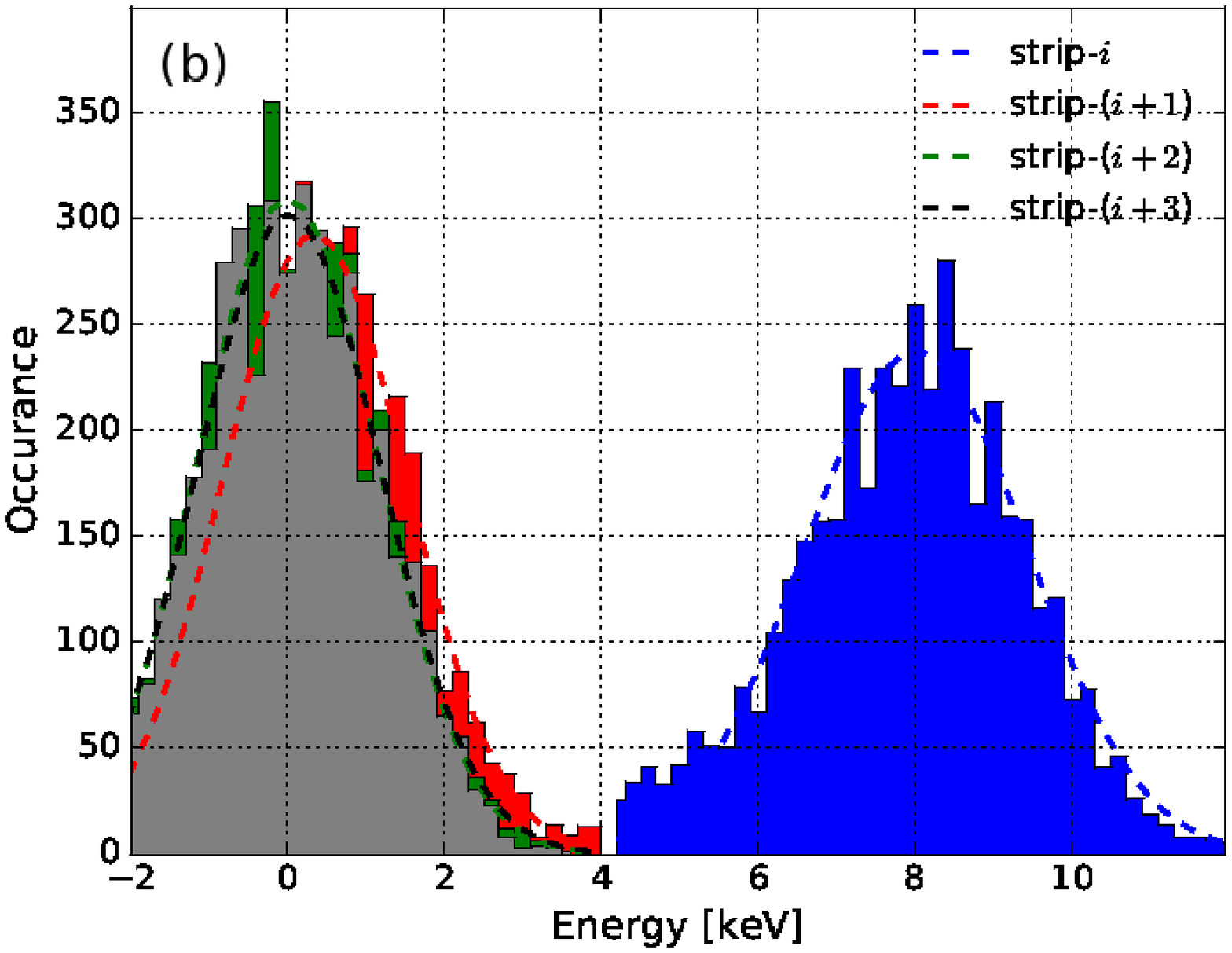}
\label{Coupling_projection}
\end{figure}

\begin{figure}
\small
\centering
\caption{Coupling factor for $k_{factor}^{1st}$, $k_{factor}^{2nd}$, $k_{factor}^{3rd}$ in HG0 and G0.}
\includegraphics[width=100mm]{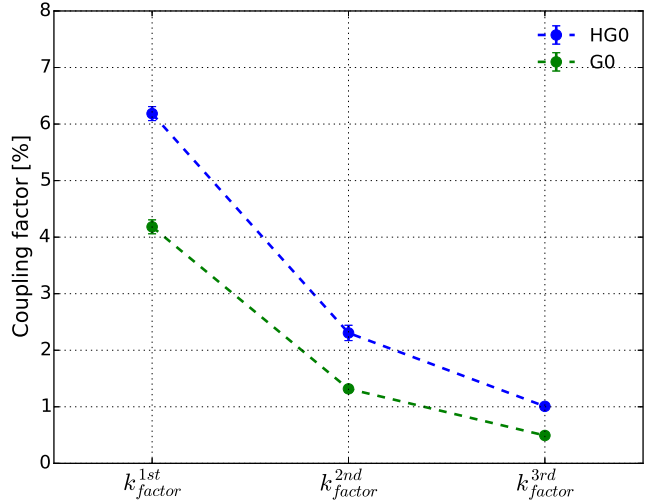}
\label{Coupling_fraction}
\end{figure}

The region with maximal occurance appearing at (0,0) in the figure refers to the 0 photon peak, and the other two regions to the single photons of \mbox{8.05 keV} in strip-$i$ and strip-($i+1$) (or strip-($i+2$), strip-($i+3$)). Taking the single photon region of strip-$i$ as the region of interest (ROI), and projecting it to the two axes, as seen in figure~\ref{Coupling_projection}(a) and (b), the energy distributions for strip-$i$, strip-($i+1$), strip-($i+2$) and strip-($i+3$) for the same entry of X-ray photons onto strip-$i$ are obtained. The mean energy/charge measured by each strip channel is obtained from Gaussian fits to each individual distribution and thus the coupling factors determined according to formula \ref{eq:k_1st}, \ref{eq:k_2nd}, \ref{eq:k_3rd}. The determined coupling factors, $k_{factor}^{1st}$, $k_{factor}^{2nd}$ and $k_{factor}^{3rd}$, as shown in figure~\ref{Coupling_fraction}, are 6.2\%, 2.3\% and 1.0\% in HG0, and 4.2\%, 1.3\% and 0.5\% in G0. Using formula~\ref{eq:Q_frac}, the fractional charges have been calculated and shown in table~\ref{Table_Qfrac}. %It is shown that strip-$i$ with X-ray impinging can only measure 84.0\% or 89.3\% of complete charge using HG0 or G0.
%The total charge is obtained by $Q^{tot} = Q^{i} + 2(Q^{i+1} + Q^{i+2} + Q^{i+3})$, assuming a symmetric coupling for strips on the left and on the right. The charge collected by the strip capacitance has not been considered due to its negligible fraction, which has been discussed previously. Thus, the the fraction of charge on collected by strip-$i$ strip-($i+1$), strip-($i+2$) and strip-($i+3$) is given by $Q^{i}/Q^{tot}$, $Q^{i+1}/Q^{tot}$, $Q^{i+2}/Q^{tot}$ and $Q^{i+3}/Q^{tot}$. The results after averaging all strips excluding the ones closing to the sensor edge have been shown in figure~\ref{Coupling_fraction}(a). In addition, the coupling factor has been calculated according to formula (\ref{eq:k_1st}, \ref{eq:k_2nd}, \ref{eq:k_3rd}) and results are shown in figure~\ref{Coupling_fraction}(b). The obtained coupling factors, $k_{factor}^{1st}$, $k_{factor}^{2nd}$ and $k_{factor}^{3rd}$, are 6.2\%, 2.3\% and 1.0\% using HG0, and 4.2\%, 1.3\% and 0.5\% using G0.

\begin{table}[htbp]
\centering
\begin{tabular}{|c|c|c|}

\hline
\textbf{Strip channel} & \textbf{Fractional charge in HG0} & \textbf{Fractional charge in G0} \\
\hline
strip-$i$ & ($84.0 \pm 0.8$)\% &  ($89.3 \pm 0.8$)\% \\
\hline
strip-($i+1$) & ($5.2 \pm 0.2$)\% & ($3.7 \pm 0.3$)\% \\
\hline
strip-($i+2$) & ($1.9 \pm 0.6$)\% & ($1.2 \pm 0.5$)\% \\
\hline
strip-($i+3$) & ($0.9 \pm 0.2$)\% & ($0.4 \pm 0.2$)\% \\
\hline

\end{tabular}
\caption{Fractional charge measured by strip-$i$, strip-($i+1$), strip-($i+2$) and strip-($i+3$).}
\label{Table_Qfrac}
\end{table}

Since the fraction of charge collected by neighbouring strips is not negligible (16.0\% in HG0 and 10.7\% in G0), this effect has to be taken into account in the detector calibration. Thus, the conversion gain obtained in Section~\ref{subsec:gain} should be corrected by dividing by a factor of 84.0\% for HG0 and 89.3\% for G0, and the noise in Section~\ref{subsec:noise} multiplying a factor of 84.0\% for HG0 and 89.3\% for G0, respectively.

%$C_{c} = \pi \epsilon_{0} l/ \textrm{ln} \left[ \dfrac{d}{2a} + \sqrt{\dfrac{d^{2}}{4a^{2}} -1} \right] $

\begin{figure}
\small
\centering
\caption{Simulation result for the inverting pre-amplifier. (a) Output voltage vs. input voltage; (b) Derived DC gain.}
\vskip 0.2in
\includegraphics[width=75mm]{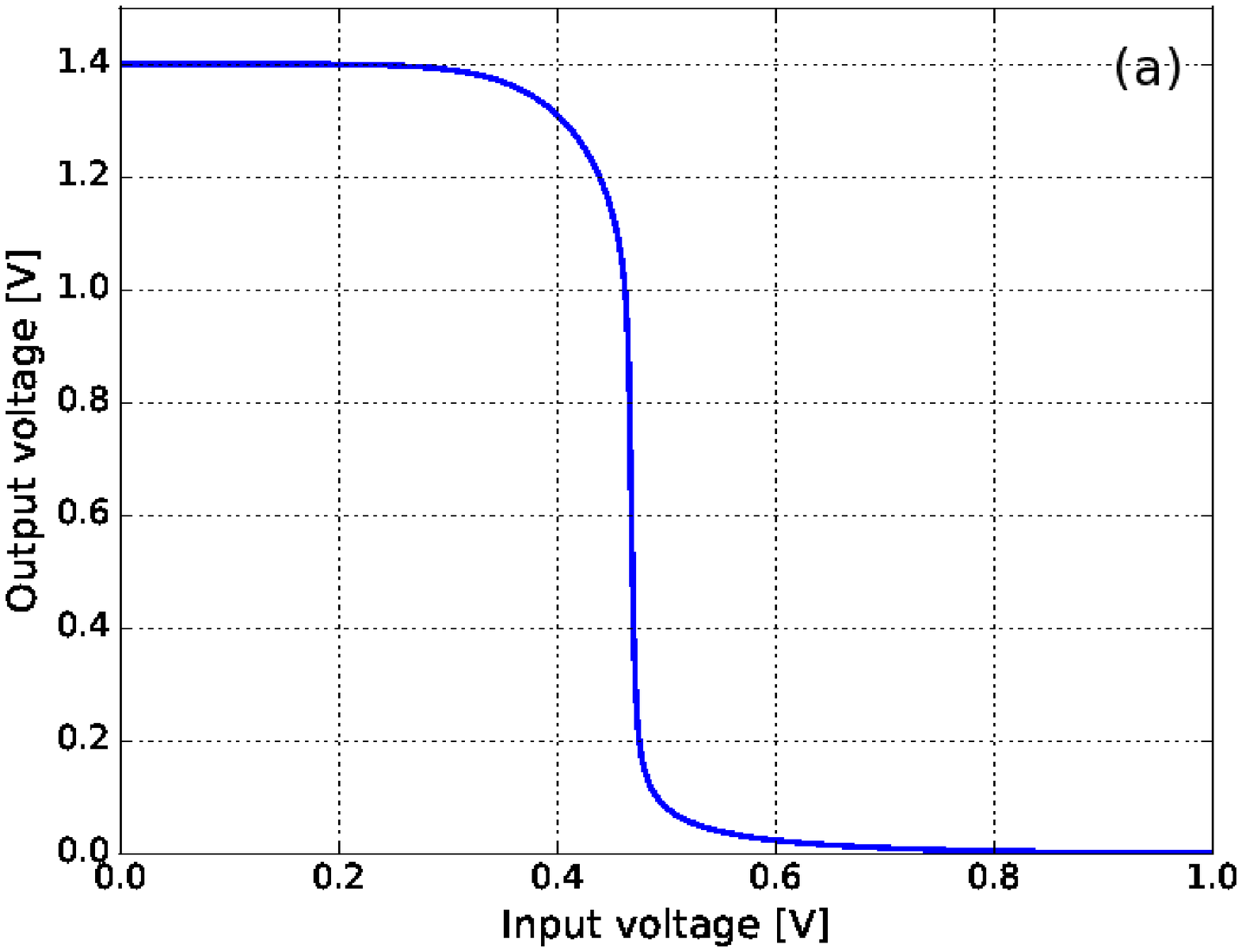}
\includegraphics[width=75mm]{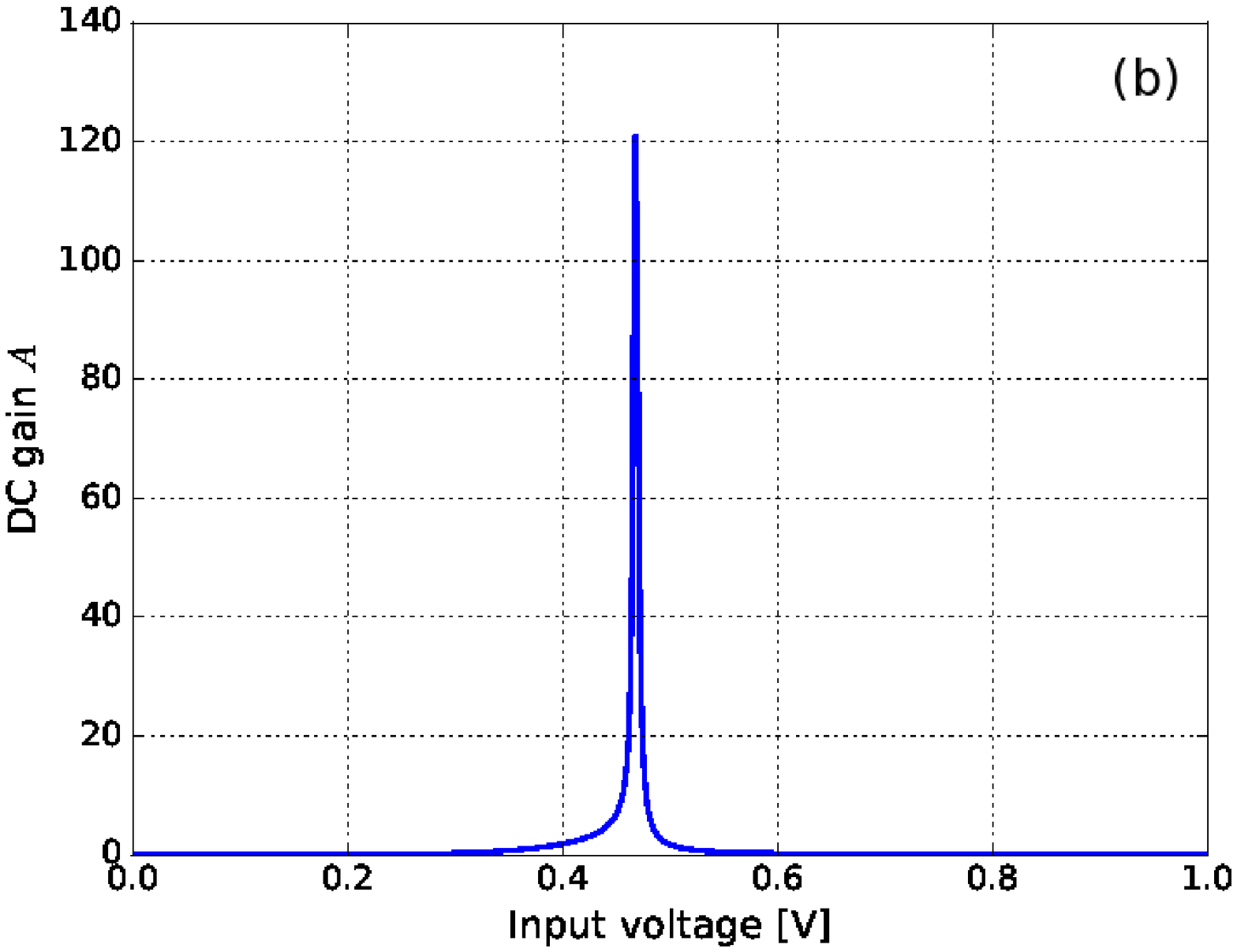}
\label{DCgain_sim}
\end{figure}

\begin{figure}
\small
\centering
\caption{Simulation results of interstrip capacitance using Synopsys TCAD for a strip sensor with \mbox{8 mm} long strips and pitch of 50 $\mu$m. (a) Simulation region of the investigated sensor with a doping of \mbox{$5 \times 10^{11}$ cm$^{-3}$} and oxide charge density of $1 \times 10^{10}$ cm$^{-2}$; (b) Interstrip capacitance as function of bias voltage.}
\includegraphics[height=60mm]{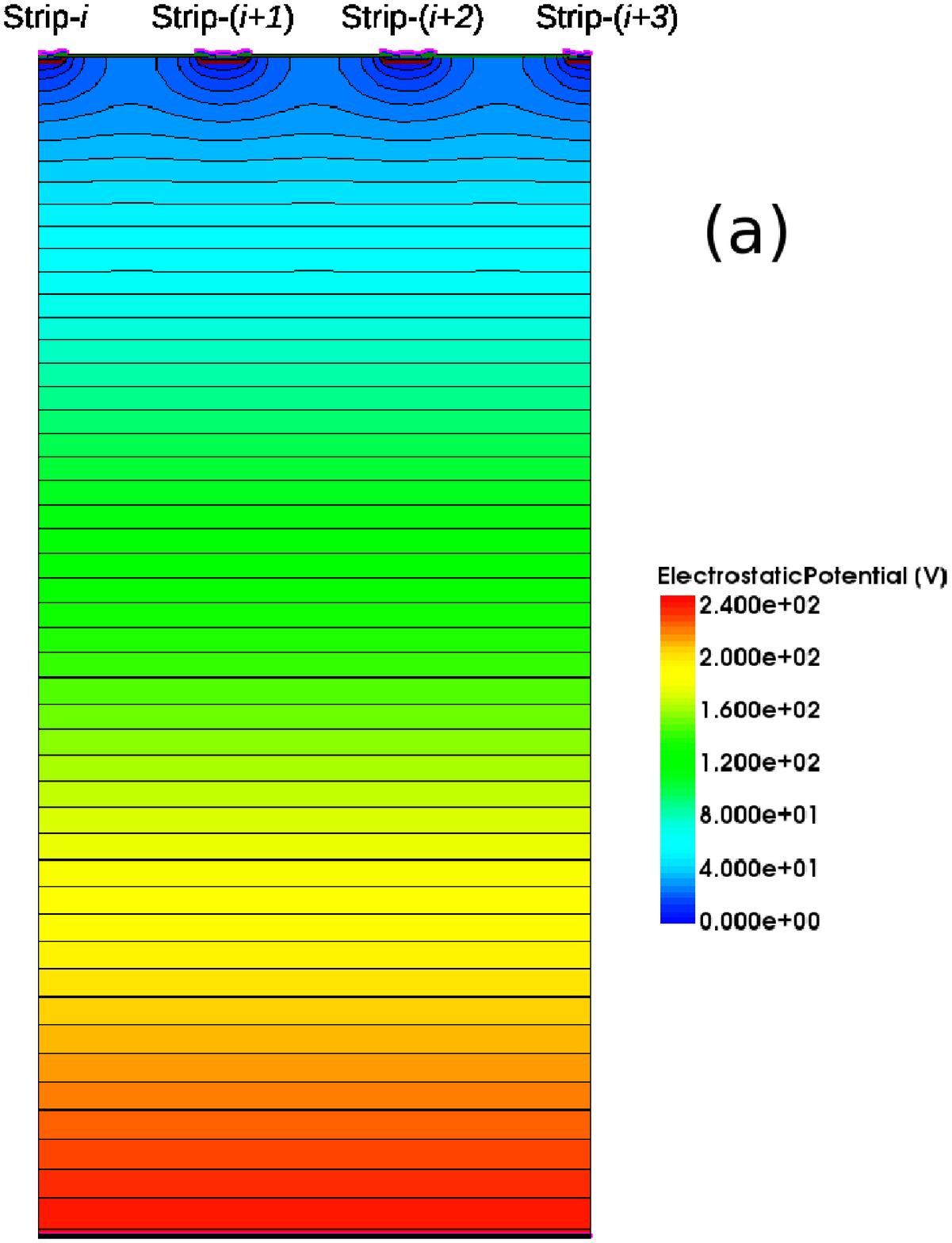}
\includegraphics[width=75mm]{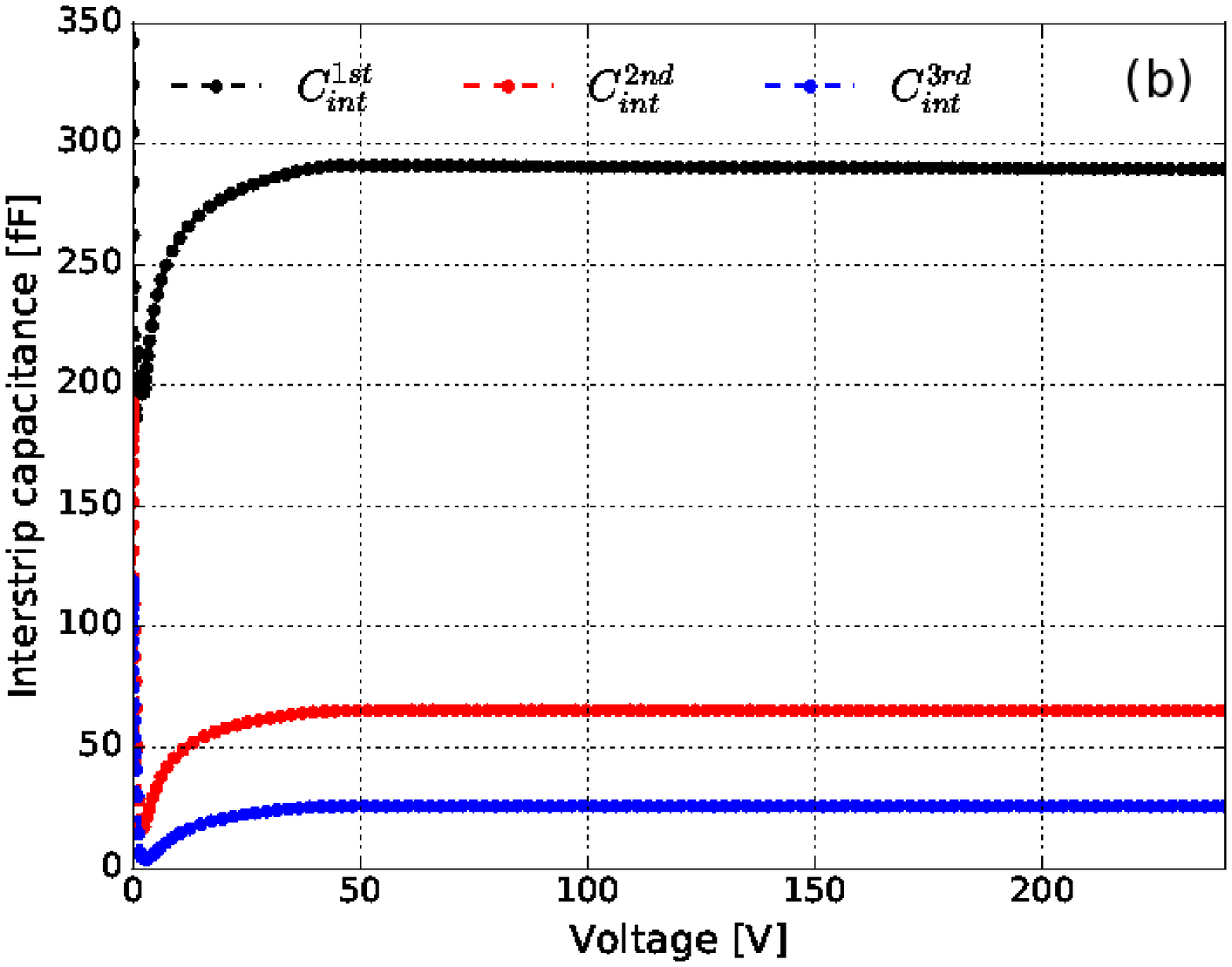}
\label{TCAD_sim}
\end{figure}

For a comprehensive understanding, the coupling factor has also been calculated theoretically. In the calculation, the DC gain of the pre-amplifer and the coupling capacitance have to be known. Since the DC gain of the pre-amplifier cannot be measured directly, it has been obtained from simulations using Cadence \cite{CADENCEweb}. Figure \ref{DCgain_sim}(a) shows the simulated output voltage of the pre-amplifier, $v_{o}^{pra}$, as function of the input voltage, $v_{i}^{pra}$. The DC gain, $A$, is derived from $A = - \Delta v_{o}^{pra} / \Delta v_{i}^{pra}$ and shown in figure~\ref{DCgain_sim}(b). $A$ is $\sim$ 121 at the working voltage of the preamplifier ($v_{o}^{pra} = v_{i}^{pra}$). The coupling capacitance, $C_{c}^{1st}$, $C_{c}^{2nd}$ and $C_{c}^{3rd}$, is mainly attributed to: Interstrip capacitance of the silicon sensor, coupling capacitance between bonding wires as well as between bonding pads of the readout channels. The interstrip capacitance is obtained from TCAD simulations \cite{TCADweb}. Figure~\ref{TCAD_sim} shows the simulated region of the strip sensor and the interstrip capacitance, $C_{int}^{1st}$, $C_{int}^{2nd}$ and $C_{int}^{3rd}$, as function of bias voltage. The values at the operation voltage of 240 V are 287.1 fF, 65.2 fF and \mbox{25.8 fF}, respectively. The simulated results agree with analytical calculations  \cite{Cerdeira1997, Paolo2010}. The coupling capacitance between bonding wires, $C_{c,wire}^{1st}$, $C_{c,wire}^{2nd}$, and $C_{c,wire}^{3rd}$, are 70.4, 46.9 and \mbox{39.2 fF} based on a theoretical calculation for pairs of parallel wires\footnote{The coupling capacitance between bonding wires is given by $C_{c,wire} = \pi \epsilon_{0} l/ \textrm{ln} \left[ d/(2a) + \sqrt{d^{2}/(4a^{2}) -1} \right] $~\cite{Lawrence1996}. $\epsilon_{0}$ is the permittivity of free space; $d$ is the distance between two parallel bonding wires with a radius of $a$ and a length of $l$. In the calculation, $a$ = 12.5 $\mu$m, $l$ = 3.5 mm, and $d$ = 50, 100 and 150 $\mu$m were used, assuming independent coupling between each pair of wires. Thus, $C_{c,wire}^{1st,2nd,3rd}$ = 70.4 fF, 46.9 fF and 39.2 fF.}; the coupling capacitance between bonding pads of strip-$i$ and strip-($i+1$) is found to be \mbox{35.4 fF}, while the capacitance between strip-$i$ and the others is negligible\footnote{The value is derived by measuring the coupling factor $k_{factor}^{1st}$ between strip-$i$ and strip-($i+1$) after removing the bonding wires of strip-($i+1$).}. Taking all the contributions into account, the coupling capacitance, $C_{c}^{1st}$, $C_{c}^{2nd}$ and $C_{c}^{3rd}$, are approximately \mbox{105.8 fF}, \mbox{46.9 fF} and \mbox{39.2 fF}, respectively. Giving the fact that the DC gain of the pre-amplifier, the feedback capacitance, and its parasitic, as well as coupling capacitance have been obtained from previous determination, the coupling factor $k_{factor}^{1st}$, $k_{factor}^{2nd}$ and $k_{factor}^{3rd}$ are 6.0\%, 1.7\% and 1.0\% in HG0, and 3.7\%, 1.1\% and 0.6\% in G0 based on theoretical calculations using formula~\ref{eq:Q_frac}. 

Table \ref{Table_spec} shows the comparison of the coupling factors obtained from measurements and theoretical calculations, and the differences are within $\sim$ 30\%. The difference can be attributed to: 1) the simple assumption $C_{c}^{1st,2nd,3rd} << (A+1)\cdot (C_{f}+C_{para})$, which neglects the charge division in-between the other channels without X-ray photons incoming, (2) the over-estimation of the DC gain of the pre-amplifier which depends on the input and output voltage of the pre-amplifier and might be different in the measurement, (3) mismatch of the feedback capacitance of the pre-amplifier in ASICs fabrication, and (4) the rough estimation of coupling capacitance between bonding wires, under the assumptions that the coupling between different pair of wires is independent and the wires are parallel and equal distance from one to another.

\begin{table}[htbp]
\centering
\begin{tabular}{|c|c|c|}

\hline
\textbf{Coupling factor} & \textbf{Measurement} & \textbf{Calculation} \\
\hline
$k_{factor}^{1st}$ (in HG0) & ($6.2 \pm 0.1$)\% & 6.0\%  \\
\hline
$k_{factor}^{2nd}$ (in HG0) & ($2.3 \pm 0.1$)\% & 1.7\% \\
\hline
$k_{factor}^{3rd}$ (in HG0) & ($1.0 \pm 0.02$)\% & 1.0\% \\
\hline
$k_{factor}^{1st}$ (in G0) & ($4.2 \pm 0.1$)\% & 3.7\% \\
\hline
$k_{factor}^{2nd}$ (in G0) & ($1.3 \pm 0.1$)\% & 1.1\% \\
\hline
$k_{factor}^{3rd}$ (in G0) & ($0.5 \pm 0.01$)\% & 0.6\% \\
\hline

\end{tabular}
\caption{Comparison of coupling factors between measurement results and theoretical calculations in HG0 and G0.}
\label{Table_comp}
\end{table}

\subsubsection{Coupling right after dynamic gain switching}
\label{sec:coupling}

%When dynamic gain switching happens, in principle, the output of the pre-amplifier can be brought to the voltage used for pre-charging the medium and low gain feedback capacitors, $C_{f,G1}$ and $C_{f,G2}$, only if the following conditions can be satisfied: 1) the capacitive coupling is negligible, 2) $C_{f,HG0}$ and $C_{f,G0}$ are small enough compared to $C_{f,G1}$ and $C_{f,G2}$, and 3) charge stored on the switches in the branches of $C_{f,G1}$ and $C_{f,G2}$ is negligible as well. However, the charge stored on $C_{f,G1}$ and $C_{f,G2}$ due to pre-charge will be re-distributed to all capacitors in the feedback loop of the pre-amplifier once gain switching occurs and to the neighbouring channels as well due to the capacitive coupling.
When dynamic gain switching happens in one strip channel, the charge stored on $C_{f,G1}$ during the pre-charge phase (and $C_{f,G2}$ if the second gain switching occurs) will be re-distributed to all capacitors in the feedback loop of the pre-amplifier and to the neighbouring channels as well due to the capacitive coupling. In this case, the charge division into neighbouring channels is also reduced due to the equivalent capacitance of the pre-amplifer of the switched strip channel increases from $(A+1)\cdot (C_{f,HG0}+C_{para})$ to $(A+1)\cdot (C_{f,HG0}+C_{f,G1}+C_{para})$, according to formula~\ref{eq:Q1st}, \ref{eq:Q2nd}, \ref{eq:Q3rd}. This causes: (a) an abrupt change of the charge in the neighbouring channels without gain switching; (b) a delay of gain switching of the neighbouring channels.
%the CDS stage will be by-passed and the output of the pre-amplifier will be brought to the voltage used for pre-charging the medium and low gain capacitors. The change of the output of the pre-amplifier results in a change of output in its neighbouring channels due to the capacitive coupling.

\begin{figure}
\small
\centering
\caption{Strip-to-strip coupling at the gain switching point using HG0. Infrared laser was injected into the center of strip-$i$. Measured ADU of strip-$i$, strip-($i+1$), strip-($i+2$) and strip-($i+3$) as function of number of 12.4 keV photons. strip-$i$ switches at $\sim$ 25 photons of 12.4 keV and its output decreases after gain switching due to the CDS stage has been by-passed.}
\vskip 0.2in
\includegraphics[width=100mm]{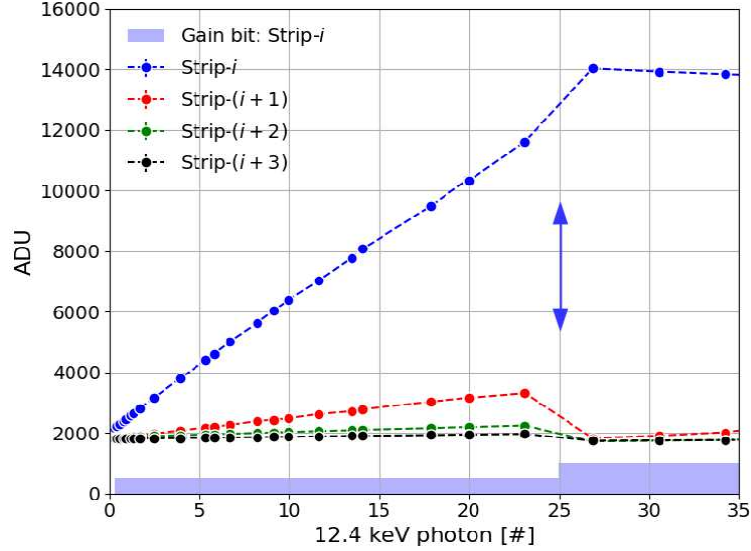}
\label{Coupling_DR_center}
\end{figure}

The first phenomena (a) is observed experimentally using fore-mentioned infrared laser injecting into the center of a strip (strip-$i$) using HG0. Figure~\ref{Coupling_DR_center} shows the output of strip-$i$ and its first, second and third neighbouring strip channels. The \textit{x}-axis refers to the number of photons measured by strip-$i$. Due to the diffusion of carriers in the silicon sensor and the size of the laser beam, a fraction of the charge generated by the laser diffuses into strip-($i+1$) and thus its outputs are higher than 6.2\% of the output of strip-$i$ expected from pure capacitive coupling. When gain switching of strip-$i$ occurs (at $\sim$ 25 photons as seen in figure~\ref{Coupling_DR_center}), a reduction in ADU is clearly visible in the neighbouring strip channels. The step corresponds to 3.3 $\times$ 12.4 keV photons for strip-($i+1$), 1.1 photons for strip-($i+2$) and 0.5 photons for strip-($i+3$). Considering up to the third neighbouring strip channels, the total change is $\sim$ 9.8 $\times$ 12.4 keV photons.

\begin{figure}
\small
\centering
\caption{The SPICE model used for simulation including a RC network of different coupling sources, pre-amplifiers and CDS of seven strip channels. The injection current at the input node of the pre-amplifier of strip-$i$ was a triangle shape of 10 ns for the rise and fall time. The injected current was ramped from 55 nA to 550 $\mu$A, corresponding to 1 to 10$^{4}$ $\times$ \mbox{12.4 keV} photons respectively. Note that the CDS and gain switching electronics of each channel is not indicated in the figure.}
\vskip 0.2in
\includegraphics[width=150mm]{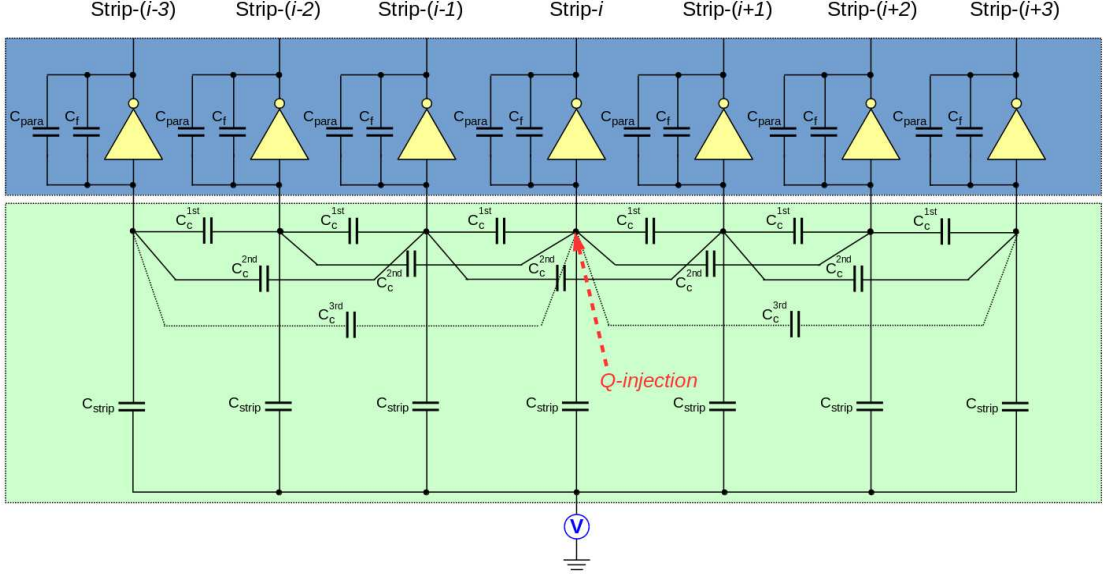}
\label{Coupling_sim_layout}
\end{figure}

The measurement results are explained by a SPICE simulation, as shown in figure~\ref{Coupling_sim_layout}: It considers a network including 7 strips, each connecting to a pre-amplifier and a CDS stage. The strip channels are coupled through interstrip capacitance as well as the coupling capacitance due to bonding wires and pads. In addition, all strips are coupled to the sensor backplane, where a bias voltage is applied. In the simulation, current was injected at the input of the pre-amplifier of strip-$i$ and the output of strip-$i$, strip-($i+1$), strip-($i+2$) and strip-($i+3$) was simulated as function of the current value. The injected current pulse was a triangle with a duration of 20 ns. The duration of the injected current pulse is longer than the pulse generated by photons in reality (usually a few ns if no "plasma effect" occurs \cite{Julian2010, Julian2010a, Julian2010b}); however, this does not influence the results. The peak values of the injected current were ramped from 55 nA to 550 $\mu$A, corresponding to 1 to 10$^{4}$ $\times$ 12.4 keV photons. %A threshold of 330 mV was set for the threshold voltage of the comparator and the feedback capacitors, $C_{f,G1}$ and $C_{f, G2}$, were pre-charged to 940 mV before currenct injection. The parasitic capacitance of 40.5 fF was added to the feedback loop of the pre-amplifier of the original design. 

\begin{figure}
\small
\centering
\caption{Simulation of the dynamic range scan showing the cross-talk after gain switching due to capacitive coupling. CDS output of strip-$i$, strip-($i+1$), strip-($i+2$) and strip-($i+3$) are shown.}
\includegraphics[width=100mm]{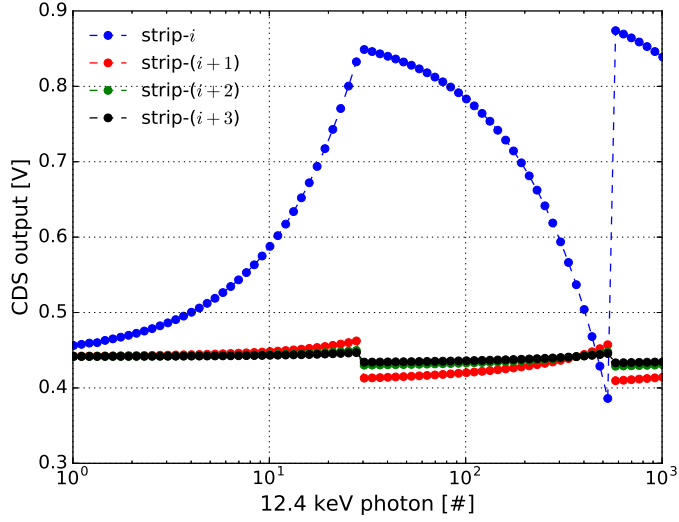}
\label{Coupling_simulation}
\end{figure}

Figure~\ref{Coupling_simulation} shows the simulation results from the output of the CDS stage as function of the number of 12.4 keV photons. For strip-$i$, the first gain switching occurs at $\sim$ 25 photons. The switching point from simulation results is consistent with the measurements. Before the gain switching of strip-$i$, the change of the CDS output for strip-($i+1$), strip-($i+2$) and strip-($i+3$) increases proportional to the output of strip-$i$ with the ratios given by the coupling factors; after the gain switching of strip-$i$, the output node of strip-$i$, which is equal to the output voltage of the pre-amplifier as the CDS is by-passed, is brought to a voltage close to the pre-charge voltage. It should be noted that the output voltage of strip-$i$ after the gain switching point is lower than the pre-charge voltage. This is mainly due to the fact that the charge pre-stored on $C_{f,G1}$ (and $C_{f,G2}$) re-distributes to $C_{f,HG0}$ and the neighbouring channels due to capacitive coupling and thus reduces the output of the switched channel. The release of the pre-stored charge into the circuit is equivalent to writing a negative charge into the input node of the pre-amplifier of strip-$i$, and thus there is a negative charge division with the neighbouring strip channels, together with the increase of the capacitive load in the feedback loop of the pre-amplifier of strip-$i$, which results in a reduction of the CDS output for strip-($i+1$), strip-($i+2$) and strip-($i+3$). The SPICE simulation qualitatively explains the measured observation.

\begin{figure}
\small
\centering
\caption{Cross-talk with infrared laser injected into the middle of the gap between two strips (strip-($i-1$) and strip-$i$). Note that the threshold voltage of the comparator in this measurement is at a lower number of photons compared to figure \ref{Coupling_DR_center}.}
\vskip 0.2in
\includegraphics[width=100mm]{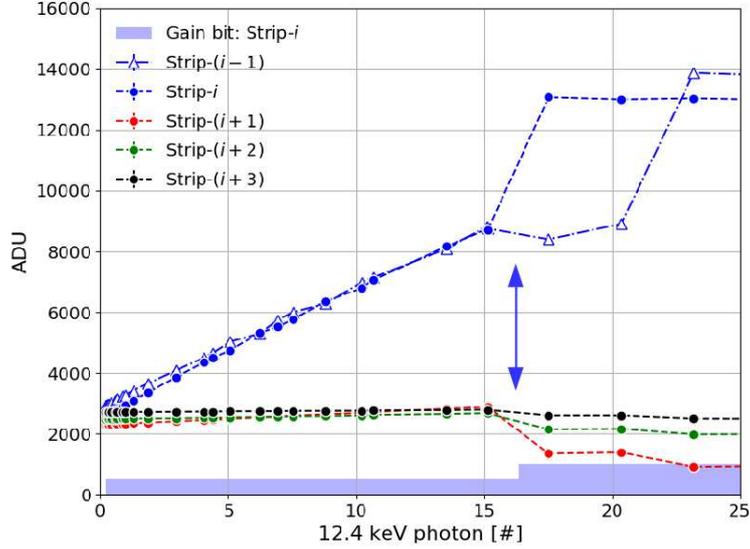}
\label{Coupling_DR_inter}
\end{figure}

The second phenomena (b) is observed through a measurement with laser injection into the middle of the gap between two strips. Figure~\ref{Coupling_DR_inter} shows the results when injecting the laser into the middle of strip-($i-1$) and strip-$i$. The results for strip-($i+1$), strip-($i+2$), strip-($i+3$) are also indicated in the figure. Due to the threshold dispersion of the channels, the gain switching point differs channel by channel. In this measurement, strip-$i$ and strip-($i-1$) receive the same amount of charge but switch at different number of photons: $\sim$ 16 photons for strip-$i$ and $\sim$ 23 photons for strip-($i+1$). Immediately after the gain switching of strip-$i$, the cross-talk reduces the signal in strip-($i-1$) by \mbox{$\sim$ 3 $\times$ 12.4 keV} photons, thus causing a further delay in gain switching of strip-($i-1$). It has been noticed that after strip-$i$ switched, the gain switching of strip-($i-1$) requires 3 photons more to switch gain, as shown in the open triangles in figure~\ref{Coupling_DR_inter}. The reason is that the charge division to strip-$i$ and strip-($i-1$) is not identical any more due to the increase of capacitive load from strip-$i$: After the gain switching of strip-$i$, its equivalent capacitance of the pre-amplifier increases from $(A+1)\cdot (C_{f,HG0}+C_{para})$ to $(A+1)\cdot (C_{f,HG0}+C_{f,G1}+C_{para})$. It means even with the same charge at the input of the pre-amplifier of strip-$i$ and strip-($i-1$), less charge flows into strip-($i-1$) due to the non-equal charge division caused by different capacitive loads. Another evidence to support the explanation is that after the gain switching of strip-$i$, the slope of strip-($i-1$) decreases, which indicates less charge collected than expected even if the charge injected into the two strips is identical. Thus a careful calibration for each channel is necessary, which requires the knowledge of the gain status of the neighbouring channels.

\begin{figure}
\small
\centering
\caption{(a) Measured ADU value of strip-$i$, strip-($i+1$), strip-($i+2$) and strip-($i+3$) as function of pre-charge voltage at the gain switching point of strip-$i$ from HG0 to G1. (b) Cross-talk in terms of number of 12.4 keV photons as function of pre-charge voltage. Note that the nominal pre-charge voltage is 940 mV.}
\vskip 0.2in
\includegraphics[width=75mm]{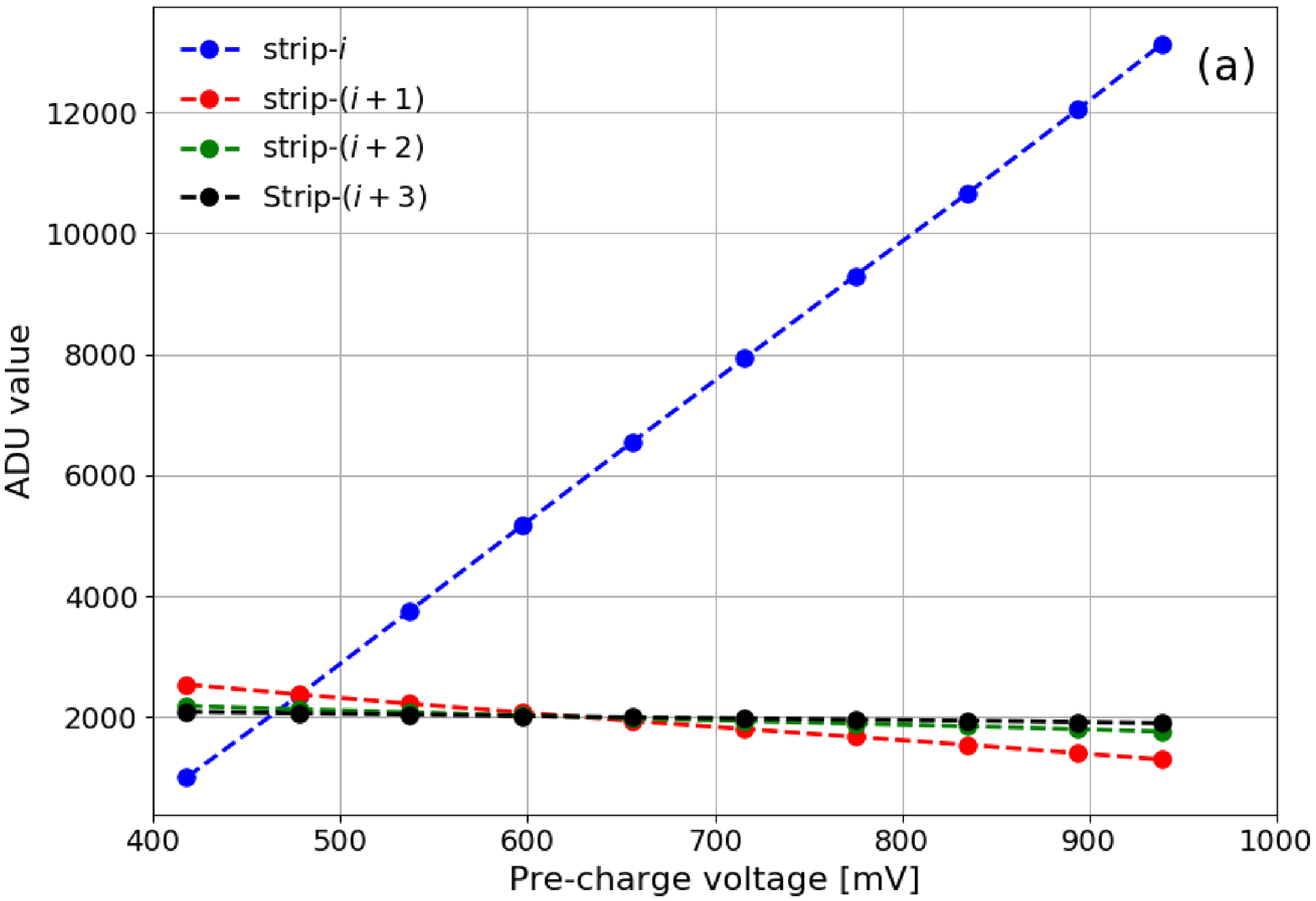}
\includegraphics[width=72.5mm]{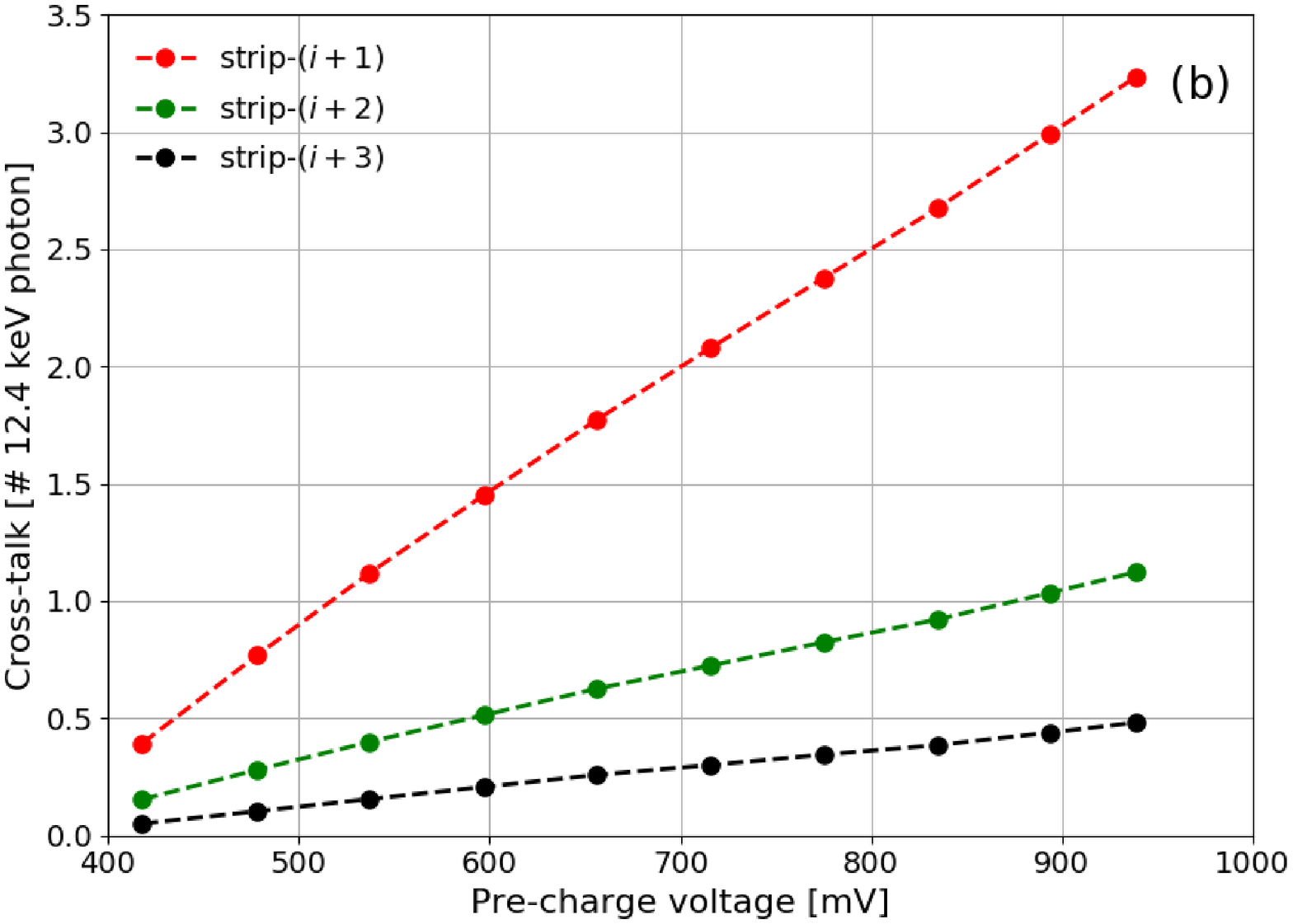}
\label{Coupling_Vdd}
\end{figure}

In addition, the cross-talk has also been investigated using different pre-charge voltages\footnote{The nominal pre-charge voltage used in the other measurements is 940 mV.}, $V_{ref,prechr}$. Figure~\ref{Coupling_Vdd}(a) shows the measured ADU value for strip-$i$, strip-($i+1$), strip-($i+2$) and strip-($i+3$) as function of $V_{ref,prechr}$, when the gain of strip-$i$ just switches from HG0 to G1. The outputs of all channels linearly depend on $V_{ref,prechr}$. The intersection point between strip-($i+1$) and strip-($i+2$) of $\sim$ 580 mV is found to be the voltage when the reset switch of the pre-amplifier is just released. Above this voltage, the measured ADU values of strip-($i+1$), strip-($i+2$) and strip-($i+3$) are below their nominal pedestal values. Figure~\ref{Coupling_Vdd}(b) is the measured cross-talk in terms of the number of 12.4 keV photons at different pre-charge voltages. With increasing the pre-charge voltage, a larger negative cross-talk is observed, due to the fact that more negative charge is pre-stored in $C_{f,G1}$ and $C_{f,G2}$. %The "charge loss" can be as high as $\sim$ 9.8 $\times$ 12.4 keV photons in the system taking into account up to the third neighbouring channel on each side of the switched channel. %For $V_{ref,prechr}$ above the voltage, negative charge is pre-stored on $C_{f,G1}$ and $C_{f,G2}$ and "charge loss" effect exists. 

The measurements indicate two ways to reduce the cross-talk: 1) Reducing the voltage used to pre-charge $C_{f,G1}$ and $C_{f,G2}$; 2) moving the working point of the pre-amplifier to a higher voltage. Both ways reduce the negative charge pre-stored in the medium and low gain capacitors, but have unacceptable side consequences: The former reduces the dynamic range; the latter increases the power of the ASIC and reduces the DC gain, which in turn increases the coupling factor before gain switching as a further drawback.

As a summary, the coupling effect can be calibrated easily before gain switching; however, after gain switching, due to the negative charge from the pre-charged feedback capacitors and the larger feedback capacitance, the charge redistributes in the readout network and a negative cross-talk is observed making detector calibration complex. Thus, reducing the coupling effect is a key task for the development of the Gotthard-II analogue front-end.

\section{Summary and discussion}

Gotthard-II is currently being under development for the XFEL.EU. It makes use of silicon strip sensor as sensing material and a dynamic gain switching ASIC to cope with the high dynamic range up to $10^{4}$ $\times$ 12.4 keV photons still keeping single photon resolution. To avoid droop effects and to achieve a compact storage of images, ADCs will be implemented in the ASIC and the digitized values will be stored in SRAM for each of the 2700 X-ray pulses and then read out during the bunch train spacing of 99.4 ms. Additional logic for digital comparisons will be designed to provide veto signals for the other pixel detectors. 

This paper puts an emphasis on the characterization of the existing analogue front-end prototypes of Gotthard: The noise, conversion gain and dynamic range are measured. Most of the results meet the specifications. The writing speed is limited by a serial resistor in the circuit which can be simply removed in the next design with an expected increase of noise as a drawback. In addition, the coupling effects have been investigated in detail. It has been found that the charge division due to capacitive coupling is not negligible in the current design; however, a careful calibration on the coupling factors for each strip channels to their neighbours make the correction for conversion gain feasible. In addition, the coupling effects have been investigated around the dynamic gain switching point: cross-talk has been observed due to the redistribution of the charge on the medium and low gain capacitors. The total cross-talk can be as high as \mbox{$\sim$ 9.8 $\times$ 12.4 keV} photons at the switching point and depends on the voltage used to pre-charge the medium and low gain capacitors. This makes the calibration of each strip complex. Thus, reducing the coupling effect is a key in further development.

Based on this study, it is known that by reducing the pre-charge voltage the cross-talk effect can be suppressed; however, this will reduce the dynamic range of the detector which is not acceptable. From the theory described in the text and the measurement results, it indicates a few ways to reduce the coupling effect: 1) Reduction of interstrip capacitance and other coupling capacitance, 2) increase of capacitance in the feedback loop of the pre-amplifier, and 3) increase of the DC gain of the pre-amplifier. For 1), the \textit{w/p} (width of the strip implant divided by the pitch) for the current design is 11 $\mu$m/50 $\mu$m = 0.22. The width of the implantation of 11 $\mu$m is close to the design limit; in addition, a further reduction of implant width cannot gain a factor of two in the reduction of the coupling factor; the coupling capacitance due to bonding wires cannot be reduced too much as well since the length of the wires is limited by the guard ring region of the silicon sensor and by the reserved safe space between the ASIC and sensor edge with high voltage. For 2), an increase of the capacitance in the feedback loop of the pre-amplifier will result in an increase of noise; in particular, since the coupling factor is inversely proportional to the feedback capacitance, a small increase in the capacitance cannot improve the coupling effect too much whereas a significant increase can result in a loss of single photon resolution due to the increase of noise. For 3), by optimizing the design of the pre-amplifier it is possible to achieve a high DC gain, and thus have the coupling effect reduced significantly. Thus, the design and optimization of a pre-amplifier with higher DC gain will be an important task in the development of the Gotthard-II analogue front-end.

\appendix

\section{List of parameters used in SPICE simulation}

The SPICE simulation started in HG0 mode and the following parameters in table~\ref{Table_SPICE} were used.

\begin{table}[htbp]
\centering
\begin{tabular}{|c|c|c|}

\hline
\textbf{Parameter} & \textbf{Value} & \textbf{Unit} \\
\hline
%$V_{th,com}$ & 330 & mV \\
%\hline
%$V_{ref,prechr}$ & 940 & mV \\
%\hline
%$V_{ref,cds}$ & 410 & mV \\
%\hline
$C_{c}^{1st}$ & 392.9 & fF \\
\hline
$C_{c}^{2nd}$ & 112.1 & fF \\
\hline
$C_{c}^{3rd}$ & 65.0 & fF \\
\hline
$C_{strip}$ & 131.7 & fF \\
\hline
$C_{para}$ & 40.5 & fF \\
\hline

\end{tabular}
\caption{Parameters used in the SPICE simulation.}
\label{Table_SPICE}
\end{table}

The values for $C_{c}^{1st}$, $C_{c}^{2nd}$ and $C_{c}^{3rd}$ are given by the sum of interstrip capacitance, coupling capacitance between bonding wires, bonding pads, as well as the metal lines at the input of the pre-amplifiers.

\section{Coupling factors derived from the dynamic range scan}

According to the measured dynamic range scan when injecting charge with a laser into the middle of two strips, the slope ratio between strip-($i+1$) and strip-$i$ before gain switching, noted as $Ratio_{i+1,i}$, the ratio between strip-($i+2$) and strip-$i$, $Ratio_{i+2,i}$, as well as the ratio between strip-($i+3$) and strip-$i$, $Ratio_{i+3,i}$, have been calculated:

\begin{equation}
\label{eq:ratio_1}
  Ratio_{i+1,i} = \dfrac{slope[\textrm{strip-}(i+1)]}{slope[\textrm{strip-}i]} = \dfrac{38.9\ [ADU/ph]}{408.9\ [ADU/ph]} = 9.5\%
\end{equation}

\begin{equation}
\label{eq:ratio_2}
  Ratio_{i+2,i} = \dfrac{slope[\textrm{strip-}(i+2)]}{slope[\textrm{strip-}i]} =  \dfrac{12.0\ [ADU/ph]}{408.9\ [ADU/ph]} = 2.9\%
\end{equation}

\begin{equation}
\label{eq:ratio_3}
  Ratio_{i+3,i} = \dfrac{slope[\textrm{strip-}(i+3)]}{slope[\textrm{strip-}i]} =  \dfrac{5.8\ [ADU/ph]}{408.9\ [ADU/ph]} = 1.4\%
\end{equation}

Assuming the coupling between strip-$i$ and strip-($i+2$) is identical to the coupling between strip-$(i-1)$ and strip-$i$, the coupling factor $k_{factor}^{1st}$ for the first neighbouring strip can be calculated by the difference of $Ratio_{i+1,i}$ and $Ratio_{i+2,i}$:

\begin{equation}
\label{eq:k_1st_cal}
  k_{factor}^{1st} = Ratio_{i+1,i} - Ratio_{i+2,i} = 6.5\%
\end{equation}

\noindent where the subtraction is made to remove the influence from strip-($i-1$). And the coupling factor $k_{factor}^{2nd}$ for the second neighbouring strip and $k_{factor}^{3rd}$ for the third neighbouring strip is given by:

\begin{equation}
\label{eq:k_2nd_cal}
  k_{factor}^{2nd} = k_{factor}^{1st} \cdot \dfrac{Ratio_{i+2,i}}{Ratio_{i+1,i}} = 6.5\% \cdot \frac{2.9\%}{9.5\%} = 2.0\%
\end{equation}

\begin{equation}
\label{eq:k_3rd_cal}
  k_{factor}^{3rd} = k_{factor}^{2nd} \cdot \dfrac{Ratio_{i+3,i}}{Ratio_{i+2,i}} = 2.0\% \cdot \frac{1.4\%}{2.9\%} = 1.0\%
\end{equation}

The derived coupling factors $k_{factor}^{1st}$, $k_{factor}^{2nd}$ and $k_{factor}^{3rd}$ agree with the extraction from low-rate X-ray measurement quite well.

%\acknowledgments

%This is the most common positions for acknowledgments. A macro is available to maintain the same layout and spelling of the heading.

%\paragraph{Note added.} This is also a good position for notes added after the paper has been written.

% We suggest to always provide author, title and journal data:
% in short all the informations that clearly identify a document.

\end{document}